\documentclass{article}

\usepackage{arxiv}
\usepackage{eurosym}
\usepackage[utf8]{inputenc} 
\usepackage[T1]{fontenc}    
\usepackage{hyperref}       
\usepackage{url}            
\usepackage{booktabs}       
\usepackage{amsfonts}       
\usepackage{nicefrac}       
\usepackage{microtype}      
\usepackage{lipsum}
\usepackage{graphicx}
\usepackage{color}
\usepackage{amssymb}
\usepackage[ruled,vlined]{algorithm2e}
\usepackage{subcaption}
\usepackage{amsmath}
\usepackage[autostyle]{csquotes}
\usepackage{lscape}

\title{Anomaly detection of energy consumption in buildings: A review, current trends and new perspectives}

\author{
  Yassine Himeur$^{1}$\thanks{Applied Energy Volume 287, 1 April 2021, 116601} , Khalida Ghanem$^{2}$, Abdullah Alsalemi$^{1}$, Faycal Bensaali$^{1}$\\
  $^{1}$ Department of Electrical Engineering\\
  Qatar University, Doha, Qatar \\
  \texttt{yassine.himeur@qu.edu.qa;a.alsalemi@qu.edu.qa;f.bensaali@qu.edu.qa} \\
	 $^{2}$ Division Telecom, Center for Development of Advanced Technologies (CDTA),\\
  Algiers Algeria\\
  \texttt{kghanem@cdta.dz} \\
   \And
 Abbes Amira \\
  Institute of Artificial Intelligence\\
  De Montfort University, Leicester, United Kingdom \\
  \texttt{abbes.amira@dmu.ac.uk} \\
}

\begin{document}
\maketitle

\begin{abstract}
Enormous amounts of data are being produced everyday by sub-meters and smart sensors installed in residential buildings. If leveraged properly, that data could assist end-users, energy producers and utility companies in detecting anomalous power consumption and understanding the causes of each anomaly. Therefore, anomaly detection could stop a minor problem becoming overwhelming. Moreover, it will aid in better decision-making to reduce wasted energy and promote sustainable and energy efficient behavior. In this regard, this paper is an in-depth review of existing anomaly detection frameworks for building energy consumption based on artificial intelligence. Specifically, an extensive survey is presented, in which a comprehensive taxonomy is introduced to classify existing algorithms based on different modules and parameters adopted, such as machine learning algorithms, feature extraction approaches, anomaly detection levels, computing platforms and application scenarios. To the best of the authors' knowledge, this is the first review article that discusses anomaly detection in building energy consumption. Moving forward, important findings along with domain-specific problems, difficulties and challenges that remain unresolved are thoroughly discussed, including the absence of: (i) precise definitions of anomalous power consumption, (ii) annotated datasets, (iii) unified metrics to assess the performance of existing solutions, (iv) platforms for reproducibility and (v) privacy-preservation. Following, insights about current research trends are discussed to widen the applications and effectiveness of the anomaly detection technology before deriving future directions attracting significant attention. This article serves as a comprehensive reference to understand the current technological progress in anomaly detection of energy consumption based on artificial intelligence.
\end{abstract}


\section{Introduction}

\begin{tabular}{|llll|}
\hline
\multicolumn{2}{|l}{ \textbf{Nomenclature}} & {\small MCU} & {\small %
microcontroller unit} \\ 
{\small ANN} & {\small artificial neural network} & {\small MNN} & {\small %
mutual k-nearest neighbor} \\ 
{\small AIN} & {\small artificial immune network} & {\small MLP} & {\small %
multi-layer perceptron} \\ 
{\small ARIMA} & {\small autoregressive integrated moving average} & {\small %
MSCRED} & {\small multi-scale convolutional recurrent encoder-} \\ 
{\small CBLOF} & {\small cluster-based local outlier factor} &  & {\small %
decoder} \\ 
{\small CNN} & {\small convolutional neural network} & {\small MSE} & 
{\small multiview stacking ensemble} \\ 
{\small DAD} & {\small deep abnormality detection} & {\small NILM} & {\small %
non-intrusive load monitoring} \\ 
{\small DAE} & {\small deep autoencoder} & {\small NTL} & {\small %
non-technical loss} \\ 
{\small DBN} & {\small deep belief network} & {\small OCL} & {\small %
one-class learning} \\ 
{\small DBSCAN} & {\small density-based spatial clustering of } & {\small %
OCSVM} & {\small one-class support vector machine} \\ 
& {\small applications with noise } & {\small OCNN} & {\small one-class
neural network} \\ 
{\small DODDS} & {\small distance-based outlier detection in } & {\small %
OCCNN} & {\small one-class convolutional neural network} \\ 
& {\small data streams} & {\small OCRF} & {\small one-class random forest}
\\ 
{\small DNN} & {\small deep neural networks} & {\small PCA} & {\small %
principal component analysis } \\ 
{\small DRED} & {\small Dutch residential energy dataset} & {\small PGBOD} & 
{\small parallel graph-based outlier detection} \\ 
{\small DRL} & {\small deep reinforcement learning} & {\small PIR} & {\small %
passive-infrared } \\ 
{\small ELM} & {\small extreme learning machines} & {\small PCSiD} & {\small %
power consumption simulated dataset} \\ 
{\small GAN} & {\small generative adversarial networks} & {\small QDA} & 
{\small quadratic discriminant analysis} \\ 
{\small GBM} & {\small gradient boosting machine} & {\small QUD} & {\small %
Qatar university dataset} \\ 
{\small GTB} & {\small gradient tree boosting} & {\small RBFNN} & {\small %
radial basis function neural network} \\ 
{\small iForest} & {\small isolated forest} & {\small RBM} & {\small %
restricted Boltzmann machine} \\ 
{\small IoT} & {\small Internet of things} & {\small RNN} & {\small %
recurrent neural network} \\ 
{\small KNN} & {\small K-nearest neighbors} & {\small ROF} & {\small %
resolution-based outlier factor} \\ 
{\small KNNG} & {\small K-nearest neighbor graphs} & {\small SCiForest} & 
{\small isolation forest with split-selection criterion \ } \\ 
{\small LDA} & {\small linear discriminant analysis} & {\small semi-SVM} & 
{\small semi-supervised support vector machine} \\ 
{\small LDCOF} & {\small local density cluster-based outlier factor} & 
{\small SLFN} & {\small single-layer feed-forward neural network} \\ 
{\small LOESS} & {\small locally estimated scatterplot smoothing} & {\small %
SVM} & {\small support vector machine} \\ 
{\small LOF} & {\small local outlier factor} & {\small SVR} & {\small %
support vector regression} \\ 
{\small LSTM} & {\small long short-term memory} & {\small STTS} & {\small %
short-term time-series} \\ 
{\small MDA} & {\small multiple discriminant analysis} & {\small VFD} & 
{\small variance fractal dimension} \\ 
{\small (EM)3} & \multicolumn{3}{l|}{\small Consumer Engagement towards
Energy saving behavior by means of Exploiting Micro Moments } \\ 
& \multicolumn{3}{l|}{\small and Mobile recommendation systems} \\ \hline
\end{tabular}

\vskip5mm

Climate change is an dangerous predicament affecting the world's population. Almost 80\% of the overall world energy is produced by fossil fuels. In addition to find green energy sources, it is of utmost importance to diminish the total energy consumption percentage \cite{Himeur2020IntelliSys}. A notable approach into achieving this objective is through informing end-users of their power usage patterns. Accordingly, consumers can improve their behavior and change their consumption habits with the aim of reducing wasted energy and contributing in the promotion of sustainable and green energy ecosystems \cite{Varlamis2020CCIS}. This is quite possible, especially if recommender systems are combined with anomaly detection modules. Therefore, personalized and contextual recommendations will be generated and transmitted the end-user to assist them in adopting a more sustainable energy use behavior \cite{REHABC2020,Sardianos2020IJIS-ERS}. In this line, governments around the world have realized the importance of energy efficiency and the major role that end-users can play to curtail the entire expenditure on energy \cite{RAU2020101983}. 

On the other side, the building sector represents a major energy consumer across the world. Specifically, buildings are responsible of more than 40\% of the overall energy generated globally, which is converted to more than 30\% of the entire worldwide CO$_{2}$ emission \cite{NGARAMBE2020109807,Himeur2020iscas}. As such, the reduction of power consumption in building environments could absolutely support the urgently-needed diminutions in the world-wide power consumption and the related environmental interests. Nevertheless, reducing power consumption in buildings is not straightforward and is a challenging task since each building requires electrical energy to operate \cite{PHAM2020121082,LUO2020102283}. 
Even though there is an increasing interest towards developing zero-energy buildings, related ideas are only in their nascence 
and are just tested in limited regions of developed countries. In this context, the potential option available currently is to promote energy awareness and optimize the operation of appliances used in buildings, giving that the latter are rigorously built to consume the amount of energy needed for their expected aims, i.e. preventing energy waste \cite{HIMEUR2020INFFUS,alsalemi2019ieeesystems}. According to recent studies, people could spend up to 80--90 \% of their time in indoor environments (and extensively some unexpected circumstances, such as the COVID-19 pandemic), which can enormously impact their energy consumption levels, especially if they show negligence and carelessness \cite{CHEN2020101688,MAGAZZINO2020115835}. 

Efficient feedback could help in reducing energy consumption in buildings and lessening CO$_{2}$ emissions. Accordingly, offering updated information and personalized recommendations to end-users and building managers is the initial stage towards setting innovative approaches to optimize energy usage \cite{BRULISAUER2020111742,Himeur2020icict}. In addition, for effective power usage, anomalous consumption behavior must be captured \cite{HIMEUR2020114877}. Therefore, via implementing energy monitoring systems and benchmarking strategies, abnormal behavior and footprints can be mitigated. Consequently, smart anomaly detection techniques for energy consumption should be formulated for identifying new forms of abnormal consumption behaviors \cite{Alsalaemi2020sca}. In buildings, an anomalous behavior of an electrical device or of the end-user could occur either because of a faulty operation of a device, end-user negligence (e.g. cold loss in a room by keeping a window open while the air conditioner is on or refrigerant leak in a fridge via maintaining the fridge door open), a theft attack, a non-technical loss, etc. \cite{Rashid8537813,HimeurCOGN2020}. An occurrence of anomalous behavior could lead to higher power consumption, longer operation-time than its normal behavior/operation-time and/or could result in a permanent malfunction of the device \cite{RASHID2019796}. 

It has been demonstrated in various research works that it should be possible to utilize artificial intelligence (AI) for detecting anomalous energy consumption behaviors either generated by end-users, appliances' failures, or other potential causes \cite{LUO2020102283,WANG2020381}. The AI community has made every possible effort during the past decade to detect abnormal power consumption accurately and swiftly. However, it is also of significant importance to detect when an appliance is not working appropriately and what are the reasons. Moreover, energy consumption events occurring during a day-off may be genuine, or harder to deal with compared to recurring events, and thus an anomaly detection algorithm might consider a recurring fault as \enquote{normal}. This makes anomaly detection in energy consumption very different form other application scenarios, e.g. intrusion detection, healthcare anomaly detection, etc. \cite{Gaur8709671}. This is because (i) the other applications are drastically different as they have acute, serious consequences if the anomaly is not detected, whereas household energy anomalies might cause extract costs and jumps of the energy bills every month, but are unlikely to be life threatening; and (ii) detecting anomalous consumption should be followed by triggering a set of tailored recommendations to help end-users adjust their energy consumption habits, replace faulty appliances, identify cyber attackers on energy infrastructures and carry on legal procedures and take other measures related to end-users' negligence (e.g. close the refrigerator door, close the doors and windows of the rooms while an air conditioner is working, etc.) \cite{Rashid8683792}. Such measures could be quite useful in different ways since they result in high energy cost savings, and could further prevent different kind of disasters (e.g. a house fire).

Efficient energy saving systems based on anomaly detection schemes need to address various issues before reaching a wider adoption. Among the challenges is how to design scalable and low cost solutions while maintaining decentralization and security. Other contemporary issues include privacy preservation, consumer anonymity, and the real-time implementation of anomaly detection based systems.
A significant effort has been put in recent years to innovate anomaly detection strategies, a large amount of projects and frameworks are ongoing, which have been described in scientific journal articles, patents, reports and industrial white papers and produced principally by the academic community and industrial partners. Moreover, various AI-based anomaly detection techniques have been the subject of new energy saving solutions. However, we assert a systemic and comprehensive review conducted based on different sources is still required to investigate the challenges, issues and future perspectives of the applicability of machine learning for anomaly detection in energy consumption. In this context, this framework strives to fill that knowledge gap via proposing, to the best of the authors' knowledge, the first, extensive and timely survey of anomaly detection of energy consumption in buildings. Explicitly, with the aim of laying the foundation for this effort, the following contributions have been proposed:
\begin{itemize}
\item  First, we present an overview of existing anomaly detection schemes in building energy consumption, in which a comprehensive taxonomy is adopted to classify them into various categories based on the nature of machine learning model used to identify the anomalies, feature extraction, detection level, computing platform, application scenario and privacy preservation. In addition, we discuss various system architectures and associated modules determining the technical properties of anomaly detection systems. A considerable part of current knowledge on anomaly detection in energy consumption arises not just from conventional academic sources (i.e. journal articles and conference proceedings), but also from industrial outputs, granted patents, and whited papers. We focus in the first part of this framework on distilling valuable information from the aforementioned sources in order to allow the readers comprehending the technical challenges of energy consumption anomaly detection. More specifically, the advantages and limitations of every category is discussed thoroughly along with its competence in different case scenarios.
\item Second, we perform a critical analysis and describe by conducting an in-depth discussion of the presented state-of-the-art. We explore current difficulties and limitations issues associated with the development and implementation of the anomaly detection systems, in addition to their market barriers.
\item Third, we describe current trends and identify new challenges concerning the enrichment of anomaly detection schemes with new applications and functionalities that could impact positively the energy consumption in buildings, among them considering additional sources of data (e.g. occupancy patterns, ambient conditions, etc.), combining other technologies (i.g. non-intrusive load monitoring (NILM)), collecting annotated datasets and using unified assessment metrics. 
\item Finally, we derive a set of future research directions that require greater emphasis with regard to four aspects, in order to (i) overcome the actual drawbacks of anomaly detection algorithms, (ii) improve the exploitation of anomaly detection solutions for better energy saving ecosystems, (iii) improve the deployment of innovative anomaly detection systems in real-world scenarios, and (iv) preserving the privacy of end-users. 
\end{itemize} 

The remainder of this paper is organized as follows. An overview of state-of-the-art anomaly detection techniques in building energy consumption is presented in Section \ref{sec2}, where an exhaustive taxonomy is proposed with regards to various aspects. Furthermore, their limitations and drawbacks are highlighted. Moving forward, critical analysis and discussion are presented in Section \ref{sec3} as a result of the conducted overview, in which difficulties, limitations and market barriers are described. Following, Section \ref{sec4} is divided into two parts, in which Section \ref{sec4-1} is reserved to describing open research challenges regarding novel applications and functionalities of anomaly detection methods. While, Section \ref{sec4-2} provides a set of insightful perspectives and emerging concepts for advancing future anomaly detection systems. Finally, Section \ref{sec5} derives relevant concluding remarks.

\section{Overview of anomaly detection methods} \label{sec2}

\subsection{Overview}
\begin{figure*}[!t]
\centering
\includegraphics[width=17cm, height=15.8cm]{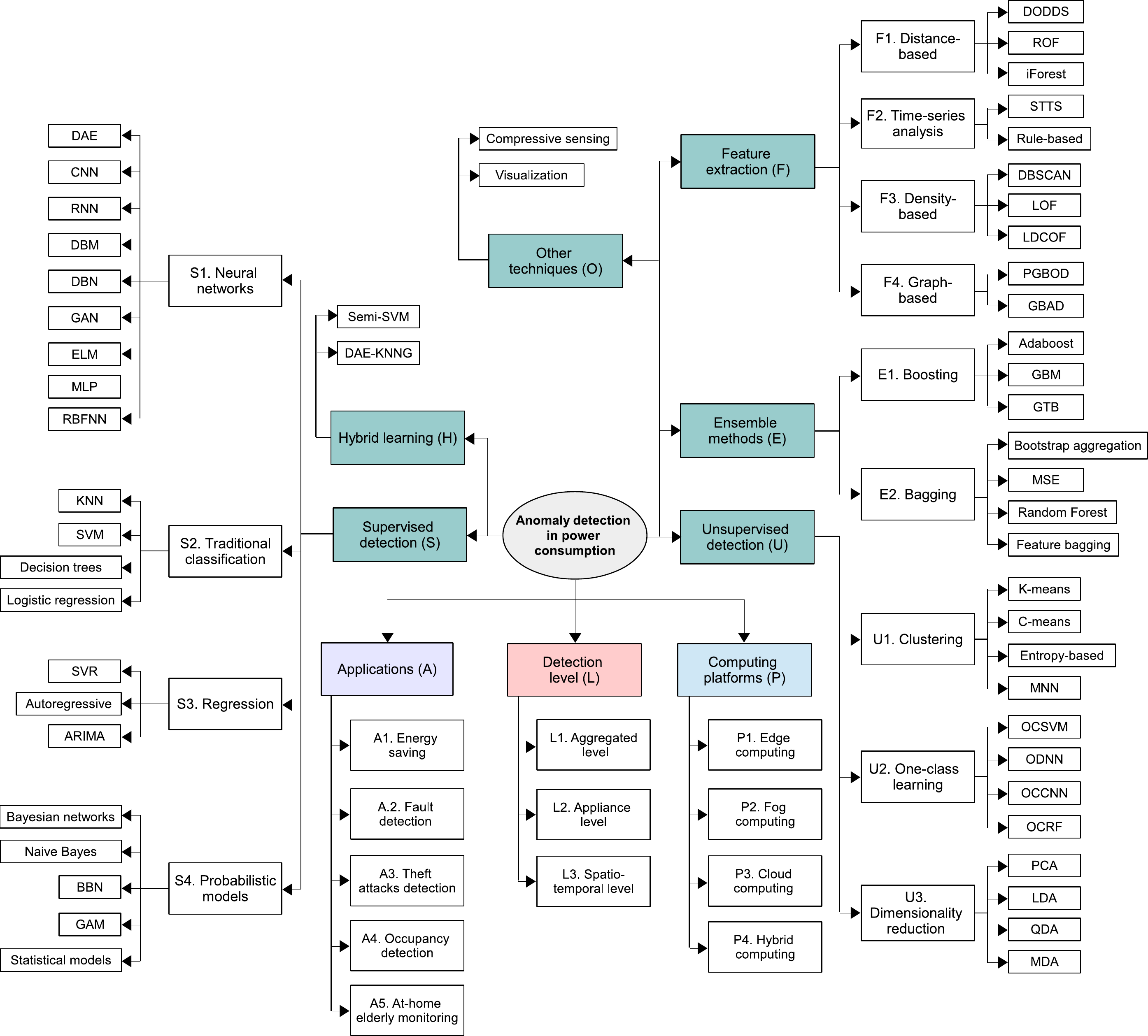}
\caption{Taxonomy of the anomaly detection schemes in energy consumption based on different aspects: i) AI algorithms, ii) application scenarios, iii) detection levels and iv) computing platforms.}
\label{taxonomy} 
\end{figure*}

This section describes existing anomaly detection methods based on the nature of implemented AI algorithms used to detect anomalies. Fig. \ref{taxonomy} illustrates the proposed taxonomy of anomaly detection techniques in building energy consumption with reference to different aspects.

\subsubsection{Unsupervised detection (U)} 
It aims at detecting formerly unknown rare consumption observations or patterns without using any a priori knowledge of these observations. Generally, this kind of detection assumes that the amount of anomaly patterns to the overall consumption data is small, i.e. less than 20\%. Since the abnormalities represent the outliers that are unknown to the consumer at the training stage, detecting anomalous consumption is reduced to the modeling of normal consumption behavior in the large majority of cases, in addition to the definition of specific measurements in this space with the aim of classifying consumption observations as abnormal or normal. Unsupervised techniques are mainly built on clustering, one-class learning and dimensionality reduction algorithms.

\vskip2mm

\noindent \textbf{U1. Clustering:} it is a machine learning scheme used to split power consumption data into various clusters and hence helps in classifying them into normal or abnormal in unlabelled datasets (even with many dimensions). This anomaly detection strategy has attracted a lot of interest in different research topics for its simplicity, such as intrusion detection in networks \cite{Kumar2016clustering}, Internet of things (IoT) \cite{doi:10.1002/ett.3647}, sensor networks \cite{Vanem2019}, suspicious behavior detection in video surveillance \cite{verma2019review}, anomalous transaction detection in banking systems \cite{AHMED2016278} and suspicious account detection in online social networks \cite{Feng2017}. In addition, clustering has the capability for learning and detecting anomalies from the consumption's time-series without explicit descriptions \cite{Pastore2020}.

Aiming at distinguishing between actual anomalies and genuine changes due to seasonal variations, the authors in \cite{Arjunan2015} propose a two-step clustering algorithm. In the first step, an anomaly score pertaining to each user is periodically evaluated by just considering his energy consumption and its variations in the past, whilst this score is adjusted in the second step by taking into account the energy consumption data in the neighborhood.  
In \cite{Rossi7844583}, the concept of \enquote{collective anomaly} is introduced, instead of the events that refer to an anomaly, to depict itemsets of events, which, depending on their patterns of appearance, might be anomalous. To achieve this, the frequent itemset mining and categorical clustering with clustering silhouette thresholding approaches were applied on smart meters data streams.
In \cite{electronics9071164} an integrated scalable framework which combines clustering and classification techniques with parallel computing capabilities is adopted, by superimposing a k-means model for separating anomalous and normal events in highly coherent clusters. Moving forward, authors in paper \cite{Izakian6608627} opt for a time-series to investigate the anomaly detection in temporal domain, subsequently to categorizing the anomalies into amplitude and shape related-ones. A unified framework is introduced to detect both type of anomalies, by employing fuzzy C-means clustering algorithm to unveil the available normal structures within the subsequences, along with a reconstruction criterion implemented to measure the dissimilarity of each subsequence to the different cluster centers. 
In \cite{Yeckle8367753}, power data are processed through the mutual k-nearest neighbor (MNN) and k-means clustering algorithms to reduce the number of measurement samples, the consumption patterns are then analyzed to detect abnormal behaviors and malicious customers. 
Finally, entropy-based methods for anomaly detection represent another clustering category, in which a little effort has been devoted to thoroughly comprehend the detection force of using entropy-based analysis, such as \cite{10.1145/1452520.1452539,e17042367}.

\vskip2mm

\noindent \textbf{U2. One-class classification:} also named one-class learning (OCL) relies on considering initial power consumption patterns to be parts of two groups, positive (normal) and negative (abnormal), then it attempts to design classification algorithms while the negative group can be either absent, poorly sampled or unclear \cite{Shi7730331}. Accordingly, OCL is a challenging classification problem that is harder to solve than conventional classification problems, which try to discriminate between data from two or more categories using training consumption data that pertain to all the groups \cite{pmlr-v80-ruff18a}. 

Different schemes have been proposed in the literature to detect anomalous consumption footprints based on OCL. In \cite{Jakkula2011}, one-class support vector machine (OCSVM) is introduced to identify the smallest hypersphere encompassing all the power observations. In \cite{chalapathy2018anomaly}, a kernel based one-class neural network (OCNN) is proposed to detect abnormal power consumption. It merges the capability of deep neural networks (DNN) to derive progressive rich representations of power signals with OCL, building a tight envelope surrounding normal power consumption patterns. In \cite{Oza8586962,ZHANG2017341}, two different approaches of one-class convolutional neural networks (OCCNN) are proposed. They share the same idea of using a zero centered Gaussian noise in the latent space as the pseudo-negative class and training the model based on the cross-entropy loss to learn an accurate representation along with the decision boundary for the considered class. 
Also, one-class random forest (OCRF) is proposed to identify abnormal consumption when labeled data are absent \cite{DESIR20133490,Ghori2020}, it is based on utilizing classifier ensemble randomization fundamentals \cite{10.1145/2133360.2133363}.

\vskip2mm

\noindent{\textbf{U3. Dimensionality reduction:}} in different machine learning applications, dimensionality reduction could be used as a classification approach with a low computational cost as it can removes irrelevant power patterns and redundancy \cite{Huang7580700}. Various techniques are explored to classify power data as normal or abnormal, such as principal component analysis (PCA), linear discriminant analysis (LDA) \cite{Valko6137278}, quadratic discriminant analysis (QDA) \cite{Naveen_2016} and multiple discriminant analysis (MDA) \cite{Brown1998}.

Despite the fact that PCA has been proposed mainly to reduce the dimensions of the original data while preserving the relationships between the data as much as possible, it has been also used as a classifier. For example, in the anomaly detection problem that is considered as a two-class classification issue, the PCA classifier estimates the principal components of both normal and abnormal classes. Following, the classifier is designed with reference to the projection of the energy patterns within the subspaces spanned by these principal components for either normal or abnormal class \cite{wu2001pca,Kudo6649443}. Moreover, PCA could also be applied for the case of multi-class anomaly detection, as it is the case with the micro-moment based anomaly detection approach described in \cite{HimeurCOGN2020}. Accordingly, the normal energy usage class has been split into three new classes while abnormal energy consumption class has been divided into two new classes. Overall, the anomaly detection problem has become a classification issue of 5 different classes. All in all, PCA is appropriate for the case in which energy observations of different categories are distributed in different spaces and directions.

In \cite{Sial2019}, the Karhunen-Loeve transform-based PCA is used to detect anomalous power consumption. It relies on estimating principal components of every consumption category and then creates a classifier via projecting power patterns on the subsets distributed by those principal components related to the two main categories (i.e. normal and abnormal). In \cite{Himeur2020IJIS-AD}, LDA is used to classify power consumption patterns by discriminating between separated sub-categories and design a model to automatically labeling power consumption patterns with reference to their corresponding categories. This has been accomplished via the use of discriminant weights to separate the hyperplanes generated by the LDA statistical learning. In \cite{Kamaraj9023207,ALHEETI2017180}, QDA that is a variant of LDA is deployed to enable a non-linear separation of power consumption patterns pertaining to both normal and abnormal ensembles. Finally, MDA is mainly used to build discriminant axes (functions) from linear combinations of the initial power consumption data. Every axis is designed to maximize the difference between normal and abnormal categories while considering them uncorrelated \cite{ALHEETI2017180,Chijoriga2011}.

\vskip2mm

\subsubsection{Supervised detection (S)} 
\begin{figure*}[!t]
\centering
\includegraphics[width=16.5cm, height=10.7cm]{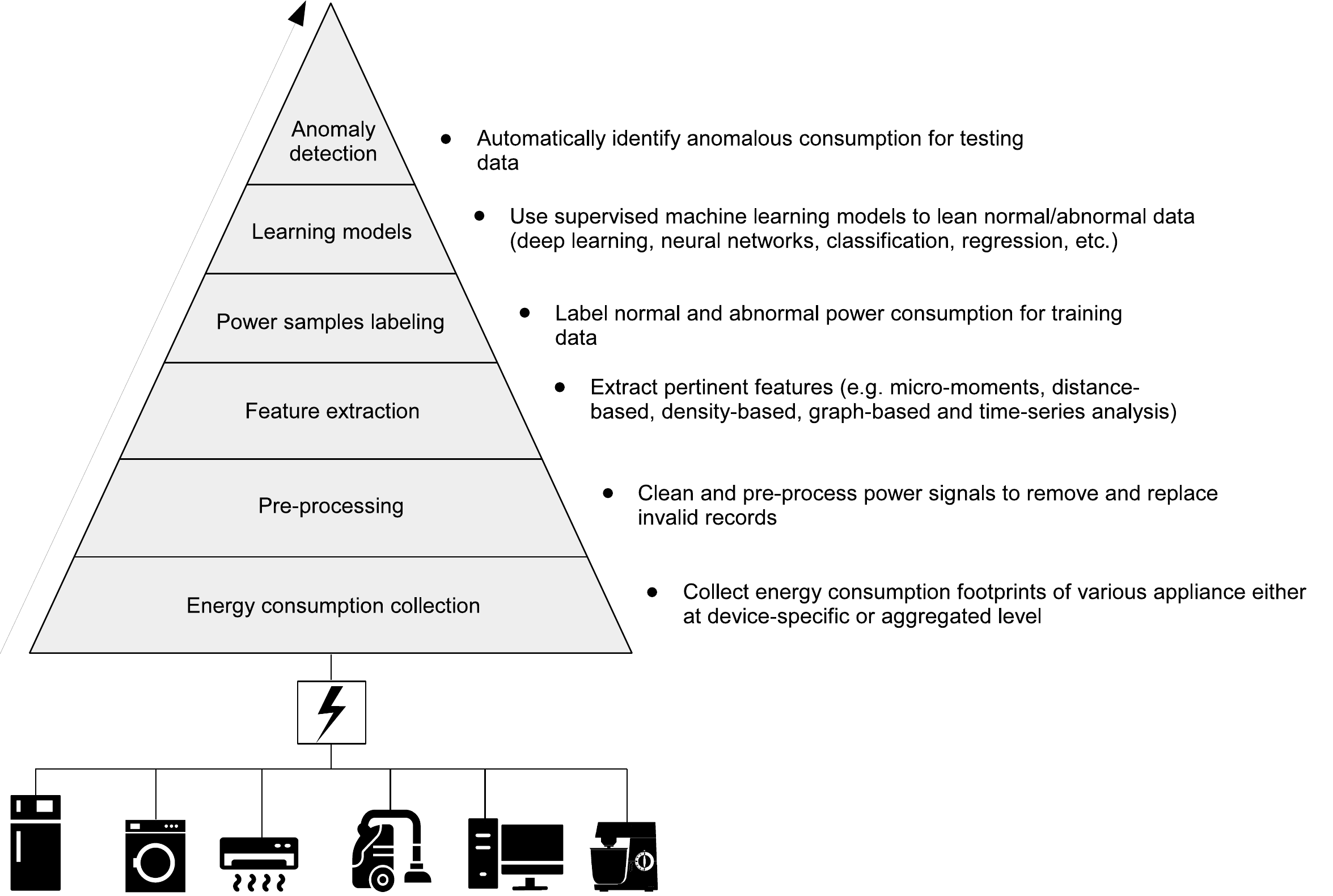}
\caption{The main steps used to perform a supervised anomaly detection of energy consumption in buildings.}
\label{SupAnoDet} 
\end{figure*}

Supervised anomaly detection in energy consumption necessitates training the machine learning classifiers (binary or multi-class) using annotated datasets, where both normal and abnormal power consumption is labeled. Although supervised anomaly detection can achieve high-accuracy identification results as demonstrated in academic frameworks, its adoption in the real world is still limited compared to unsupervised methods, due to the absence of power consumption annotated datasets. Fig. \ref{SupAnoDet} illustrates the main steps to conduct a supervised anomaly detection approach.

\vskip2mm

\noindent \textbf{S1. Neural networks:} refer to using deep learning or conventional artificial neural networks (ANN) to detect normal and abnormal consumption patterns. Currently, deep abnormality learning (DAD) has been used in various research topics, such as detecting fraudulent health-care transactions \cite{chalapathy2018anomaly}, identifying abnormalities in video streaming \cite{jimaging4020036} and detecting credit card frauds \cite{10.1145/3336191.3371876}. However, the performance of a deep learning based solution could be sub-optimal in some cases owing to the imbalance property of power consumption datasets (i.e. power consumption patterns are not uniformly distributed over the normal and abnormal categories).

In \cite{Weng8574884,Wang8725704}, the autoencoder and long short-term memory (LSTM) neural networks are merged to identify abnormalities in unbalanced and temporally correlated power consumption datasets. Similarly, in \cite{Pereira8614232}, the authors detect anomalies in time-series power footprints using a variational recurrent autoencoder. Moving forward, Yuan et Jia \cite{Yuan7415819} use stacked sparse autoencoder for extracting high-level representations from large-scale power consumption datasets gleaned using and IoT-based metering network. Next, they utilize softmax in the classification stage to capture the consumption anomalies before sending notifications and alerts to end-users using web applications. Similarly, in \cite{himeur2020detection} the autoencoder and micro-moment analysis are used to detect abnormal energy usage.

On the other side, convolutional neural networks (CNN) have demonstrated its effectiveness in different research applications, and it has superior performance in comparison with artificial neural network (ANN) algorithms for detecting abnormalities in time-series data \cite{s19112451}.
In \cite{10.1155/2019/4136874}, the author opted for combining CNN and random forest to track energy consumption anomalies due to energy theft attacks and thereby helping energy providers to remedy the issues related to irregular energy usage and inefficient electricity inspection. Similarly, Zheng et al. \cite{Zheng8233155} propose a CNN-based solution, which helps mainly in identifying the non-periodicity of energy theft and periodicity of normal energy consumption using 2D representations of power consumption signals. Using the same idea, a CNN is developed in \cite{Tang2019wiley} via representing time-series time/frequency energy consumption signals in 2D space and then learning anomaly features using convolution. Moving forward, in \cite{Zhang2018AAAI}, multi-scale convolutional recurrent encoder-decoder (MSCRED) is deployed to analyze multivariate time-series observations and detect abnormalities.
In \cite{Alrawashdeh7838144}, a restricted Boltzmann machine (RBM) along with a deep belief network (DBN) are merged to construct a DNN-based abnormality detection framework. Explicitly, a dimensionality reduction task is performed at the two first RBM layers before being fed into a fine tuning layer including a classifier to separate anomalies from normal data. 

Furthermore, looking for innovative deep learning solutions to deal with the unbalanced property of anomaly detection datasets, generative adversarial networks (GAN) are employed. It can model complex and high-dimensional data of different types, including images \cite{Choi9070362}, time-series \cite{Sun8931714,Dan2019} and cyber security \cite{HUANG2020102177}. Unfortunately, its utilization to detect anomalous power consumption in buildings is still very limited \cite{en13010130}.

Recurrent neural network (RNN) is very competent in analyzing time-series data and enables to exhibiting temporal dynamic behaviors \cite{Bontemps2016}. It has been used to predict the anomalies occurring during energy usage and distinguish them from deviations emerging from seasonality, weather and holiday dependencies \cite{daSilva8702152,Wang8725704}. For instance, in \cite{Hollingsworth8621948}, an RNN based anomaly detection system is designed, which can remove seasonality and trends from power consumption patterns, resulting in a better capture of the real abnormalities.
In \cite{Fenza8604042}, the authors concentrate on elaborating an abnormality detection scheme having the ability to face the concept drift, due to family structure changes (e.g. a household turned to a second family residence). To that end, an LSTM based RNN model is developed to profiling and forecasting end-users' consumption behaviors using their recent/past consumption data. 
In \cite{Chahla2019}, abnormal days illustrate suspicious consumption rates are identified using a hybrid learning model based on RNN and K-means. Similarly in \cite{XU2020}, a hybrid model using RNN and quantile regression is introduced to predict and detect anomalous power consumption.

\begin{figure*}[!t]
\centering
\includegraphics[width=1\columnwidth]{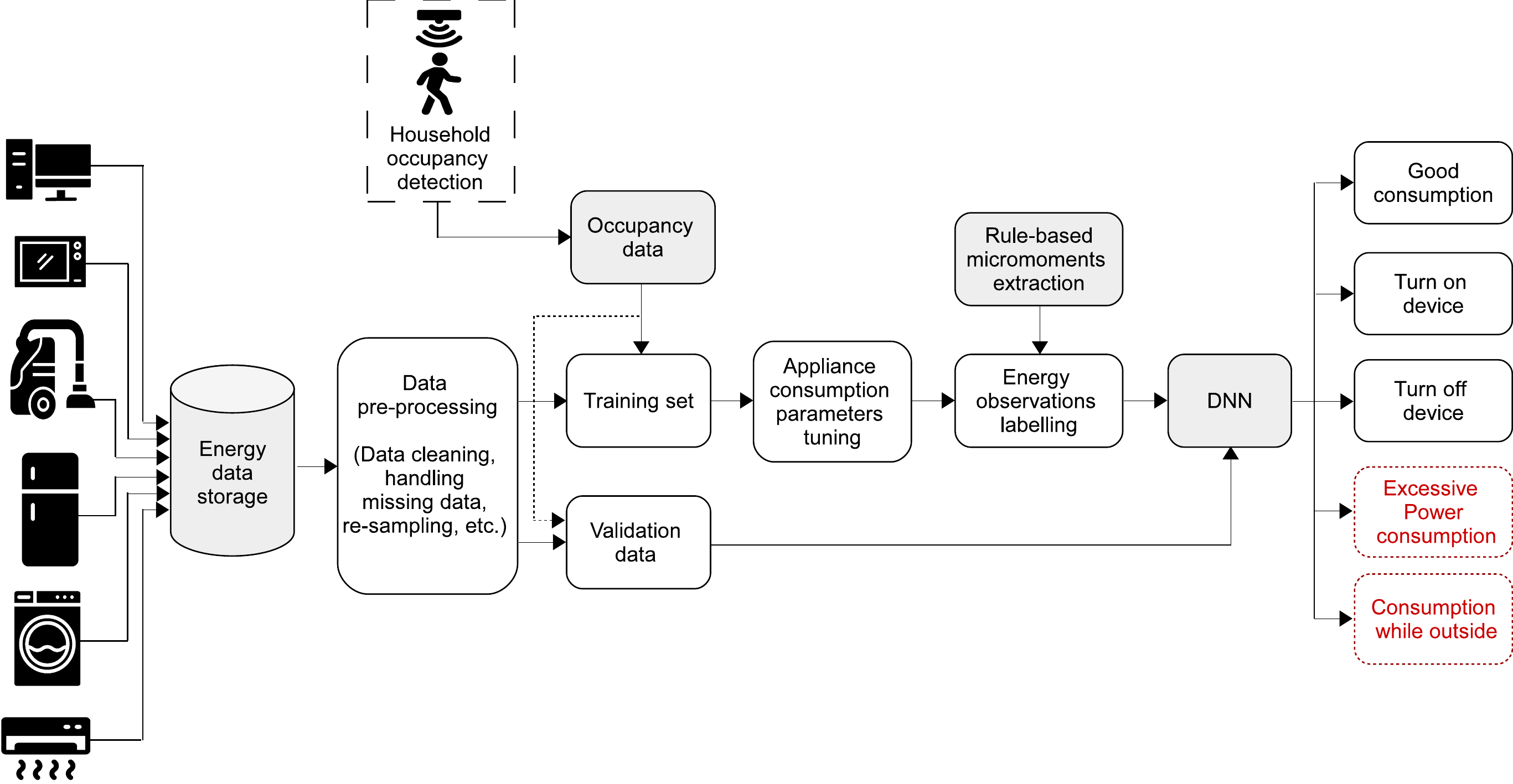}
\caption{Flowchart of a supervised anomaly detection scheme based on DNN used to detect two different anomalies, i.e. excessive consumption and consumption while outside.}
\label{Flowchart} 
\end{figure*}

In order to provide the reader with more details on the use of deep learning for anomaly detection in energy consumption, Fig. \ref{Flowchart} illustrates a flowchart of a supervised anomaly detection scheme proposed in the (EM)$^3$ project, which is performed using a DNN model \cite{HimeurCOGN2020}. In this framework, power consumption data of various appliances and occupancy patterns are gleaned using sub-meters and smart sensors. Next, collected data are labeled using a micro-moment paradigm, in which consumption footprints are divided into five consumption categories. Following, a DNN model is designed and trained using the labeled dataset before testing it on new recorded, unlabeled data in the test stage.  

\vskip2mm

On the other hand, using ANN for anomaly detection in energy consumption is mainly supported by its capability to learn and generalize from past consumption data to identify normal and abnormal behavior \cite{Chen7176083}. In addition, ANN could help in solving the anomaly detection issue when recorded data is noisy due to various reasons, e.g. noise generated during data transmission or from electrical appliances connected to the smart grid \cite{SANTOLAMAZZA20181131}. 
In \cite{Ghanbari7771715}, the identification of power consumption anomaly is handled by resorting to a multi-stage ANN-based solution. This latter incorporates a discrete wavelet transform to obtain the required features, a variance fractal dimension (VFD) operation applied on those features, an ANN scheme which exploits the VFD output to perform the training, and finally a threshold-based detection of the anomalous power consumption pattern. 
The work in \cite{WANG2020114145} proposes a residential framework comprising a dual hybrid one-step-ahead load predictor and a rule-engine-based energy consumption abnormality detector. In order to attain a high anomaly detection precision in linear and nonlinear regression, the predictor merges the benefits of ANN and autoregressive integrated moving average (ARIMA) model.  

Moreover, the consumption anomalies are tracked through the use of multi-layer perceptron (MLP) and classification techniques in \cite{Mulongo2020}. Similarly in \cite{Efferen8072036}, with the aim of predicting malicious behavior in unbalanced data, an MLP-based solution is efficiently tested on two different datasets to carry out a flow-based control which preserves the end-users' privacy. 
In the same direction, the continuous and fine-grained monitoring of energy consumption in industrial buildings is discussed in \cite{s19245370} in order to preserve reliable operation. Explicitly, an MLP-based anomaly detection scheme is targeted via detecting sensor data abnormalities in a pharma packaging system. Moreover, intrusion detection that can be applied in energy theft tracking, is investigated in \cite{Zeng2010} by combining artificial immune network (AIN) and cosine radial basis function neural network (RBFNN), wherein firstly multiple-granularities version of the former is supported to reveal the candidate hidden neurons, and subsequently, the latter is trained based on gradient descent learning process. In addition, different power consumption anomaly detection frameworks are introduced based on extreme learning machines (ELM) \cite{Janakiraman7727444,bose2018adepos}. Specifically, ELM is built upon a single-layer feed-forward neural network (SLFN) for classifying the normal and abnormal classes \cite{Imamverdiyev7991732}.

\vskip2mm

\noindent \textbf{S2. Regression:} refers to identifying the relationship between two or more power variable classes in order to produce an ensemble of model parameters to predict the generation of abnormal power observations. In this context, the production of anomalous power consumption patterns can be predicted based on other collected abnormal footprints. Various regression models have been introduced in the literature to identify abnormalities in building energy consumption, including linear regression, support vector regression (SVR), auto-regressive models, regression trees and regression fitting \cite{KROMANIS2013486}.
The authors in \cite{Zhang6039858} propose to adopt linear regression-based approaches to determine the anomalous periods for individual premises, and clear them from the premise data, such that to provide precise assessments of energy consumption patterns.
In the same direction, a model to find abnormal energy consumption patterns is designed in \cite{Fahim8587597} by analyzing the smart meters temporal data streams. Specifically, to perform the prediction and map the non-linearity of data, support vector regression with radial basis function is retained and correspondingly evaluates the disparity between the actual and the expected energy consumption. 

Because of the large quantity of stored smart meter data, anomaly detection with such information brought the big data issue into focus, particularly with the scarcity of adequate and efficient real time anomaly detection systems capable of handling this huge amount of data. In order to remedy this and facilitate energy-related decision-makings, the studies in \cite{LIU201834,info8040151} depict a scalable architecture merging an autoregressive prediction-based detection method,  with a new lambda scheme to iteratively upgrade the model along with real time anomaly detection. 
The work in \cite{CHOU2014400} targets the reduction of  anomalous consumption by presenting a new scheme which enabled the identification of anomalous power consumption within large sets of data. It follows a two-stage processing, namely prediction and then anomaly detection, where, by the aid of a hybrid neural network ARIMA model of daily consumption, daily real-time consumption is first predicted in the former step, whereas a two-sigma rule was adopted to localize the anomalies via the evaluation of the mismatch between real and predicted consumption.
The framework in \cite{s19112451} address the anomaly recognition in streaming large scale data, which is a typical occurrence scenario in deployed sensors. In this scope, both statistical (i.e. ARIMA) and CNN based approaches were integrated in a residual way, such that the fusion was shown to compensate the weaknesses of each of them and consolidate their strengths. In \cite{s19245370}, a data-driven approach was pursued since no cyclicity pattern was noted on the observed data. From comparing three different regressors (i.e. regression tree, random forest, and MLP) in the prediction phase, the authors highlighted the advantages of the regression trees and random forests residing in the training time  efficiency and model replicability ease.

\vskip2mm

\noindent \textbf{S3. Probabilistic models:} are among the most important machine learning tools, they have been instituted as an effective idiom for describing the real-world problems of anomaly detection in energy consumption using randomly generated variables, such as building models represented by probabilistic relationships \cite{Bacciu2015,Ethan2021}. 
The anomaly profiles of time-series patterns are identified using Bayesian maximum likelihood models for clean data \cite{AKOUEMO2016948} and noisy data \cite{Ethan2021}, while Bayesian network models are implemented to detect abnormalities categorical and mixed based power consumption data in \cite{Rashidi2011,Saqaeeyan2020}. 
In \cite{Jakkula5673788,Liu2016lambda}, statistical algorithms are deployed to identify the anomalies via the identification of extremes based on the standard deviation, while in \cite{Jakkula5673788}, the authors use both statistical models and clustering schemes to detect power consumption anomalies. In \cite{HOCK2020100290,Coma-Puig7796897}, naive Bayes algorithms are proposed to detect the abnormalities generated by electricity theft attacks. Similarly in \cite{Janakiram1665221}, Janakiram et al. deploy a belief Bayesian network to capture the conditional dependencies between data and then identify the anomalies. In \cite{Chen10.1109/ICPR}, a statistical prediction approach based on a generalized additive model is introduced to timely detect abnormal energy consumption behavior.  

\vskip2mm

\noindent{\textbf{S4. Traditional classification:}} stands for models that rely on detecting to which power consumption category (sub-population) a new power consumption sample pertains, with reference to a training ensemble of consumption footprints that have labels of both normal and anomalous consumptions. K-nearest neighbors (KNN), support vector machine (SVM), decision tree and logistic regression are the well-known conventional classification algorithms, they have been widely deployed in the state-of-the art of the energy-based applications or other research topics.

In \cite{Jakkula5673788,Sial2019}, KNN based heuristics are proposed to detect abnormal power consumption, while in \cite{Mulongo2020}, the authors investigate the performance of KNN against other machine learning classifiers to identify abnormal power observations. In \cite{Depuru5772466,Korba8807832}, SVM is deployed to detect abnormalities due to energy theft attacks. In the same direction, in \cite{Nagi4766403}, a genetic SVM model is proposed to detect abnormal consumption data and suspicious customers, in which a genetic algorithm is combined with SVM. While in \cite{Zhang8974989}, Zhang et al. fuse SVM and particle swarm optimization for detecting abnormal power consumption in advanced metering infrastructures.
On the other side, in \cite{Cody7424479}, a decision tree based solution is introduced to learn energy consumption anomalies triggered by fraud energy usage. Similarly in \cite{Reif4761796}, an improved decision tree model is developed to detect anomalous consumption data using densities of the anomaly and normal classes. Moving forward, in \cite{s19245370}, a decision tree regressor is presented to detect abnormal power consumption using sensor data, while in \cite{Mulongo2020}, the anomalies are detected using logistic regression.  

\subsubsection{Ensemble methods (E)}
As it is demonstrated in various frameworks \cite{HimeurCOGN2020,Xu2019eurasip}, none of the anomaly detection schemes could perfectly identify all abnormalities through low-dimensional subspaces because of the complexity of power consumption data and other factors influencing power usage over hourly, daily, weekly, monthly or yearly scales. Accordingly, the use of ensemble learning can solve some related issues, where the initial set of power observations is split to multiple subsets and various models are applied simultaneously on these subsets to derive the potential abnormalities. Following, anomaly identification scores are either summarized or the most suitable one is selected to produce final score.

\vskip2mm

\noindent \textbf{E1. Boosting:} it is a set of meta-algorithms used to principally reduce bias and variance of unsupervised learning, in which weak classifiers (learners) are converted into strong ones. Generally, they are structured in a sequential form. A weak classifier refers to the case where a slight correlation can be achieved with the true classification \cite{Bahri2011}. Different boosting schemes are proposed in literature to detect anomalies, among them bootstrap, gradient boosting machine (GBM) and gradient tree boosting (GTB).

In \cite{Zhang8467985}, Zhang et al. use a bootstrap strategy to conduct an unlabeled learning process for detecting anomalies in energy data in multi-feature data. In \cite{TOUZANI20181533}, a GBM based anomaly detection is introduced to model power usage of commercial buildings. In the same manner, in \cite{Tama2019}, a grid search is deployed to capture the best parameter configuration of a GBM based anomaly detection. While in \cite{Albiero2019}, the authors predict energy frauds though the identification of power consumption anomalies using a GBM based scheme. In \cite{Kim7463810}, a GTB based anomaly detection is investigated along with other data mining techniques using power consumption pricing data.

\vskip2mm

\noindent \textbf{E2. Bagging:} also called bootstrap-aggregating, it is a set meta-algorithms developed for improving the accuracy and stability of several weak classifiers. Bagging differs from boosting by the fact that the weak learners are structured in a parallel form \cite{Gaikwad7155853}. Moreover, distinct detection schemes can be applied on each sub-ensemble before aggregating their results as demonstrated in \cite{Nguyen2010}. Random forests, bootstrap aggregation and their variations are the well-known bagging based ensemble learning methods used for anomaly detection. For example, in \cite{ARAYA2017191}, Araya et al. propose a bootstrap aggregation based abnormality detection scheme, which helps in conducting an ensemble learning to identify energy consumption anomalies. In \cite{Wun8974857}, an isolation forest with split-selection criterion (SCiForest) algorithm is introduced  to check if the end-user's electricity consumption is anomalous or normal. In \cite{10.1155/2019/4136874}, non-technical losses (NTLs) occurring in the energy networks are detected using a random forest scheme. This is mainly conducted through sensing anomalous power consumption and learning consumption differences for different periods (i.e. hours and days). 

In \cite{Primartha8285847}, a random forest classifier is deployed to detect anomalies while respecting the performance measure related to the accuracy and false alarm rates.
In \cite{8291600Ouyang}, a multiview stacking ensemble (MSE) technique is proposed to learn energy consumption anomalies collected using different IoT sensors in industrial environments.
In \cite{Xu2019eurasip}, an anomaly detection scheme based on feature bagging is introduced. It relies on training several classifiers on different feature sub-ensembles extracted from a main high-dimensional feature set and therefore combining the classifiers' results into a unique decision. 
In \cite{Lazarevic2005}, after deriving various feature sub-ensembles randomly from the initial feature, anomalies are identified and the performance is estimated in each sub-ensemble before fusing them to come out with the final output. 

\subsubsection{Feature extraction (F)}
This part mainly discusses how feature extraction scheme can help to boost the performance of anomaly detection methods via: (i) representing the power consumption observations in novel spaces (e.g. high-dimensional spaces); (ii) utilizing appropriate measures and functions (e.g. distance, density) to discriminate between normal and abnormal consumption; and (iii) representing the consumption flowchart using new representation structures (e.g. graph-based representation) \cite{Araya7727242}.

\vskip2mm

\noindent \textbf{F1. Distance-based:} refers to detecting abnormal consumption patterns by judging each pattern based on its distance to its neighboring samples. Explicitly, normal consumption observations generally possess a dense neighborhood while anomalous consumption footprints are far form their neighboring points (i.e. show a sparse structure). Various frameworks have been proposed to resolve the issue of distance-based anomaly detection for energy consumption, where unsupervised learning methods are usually adopted without having any distributive presumptions on recorded consumption data. In this regard, in \cite{Gu2019stat}, a distance-based anomaly detection is proposed via analyzing the theoretical properties of the nearest neighbors of each power observation. Explicitly, anomalous patterns are then detected with reference to a global quantity named distance-to-measure. Also in \cite{Zhang8636303}, power anomalies in smart grid are detected using a multi-feature fusion that is based on Euclidean distance and a fuzzy classification approach. In \cite{Yijia7581646}, the authors use a cosine similarity approach to estimate similarity distance between power consumption observations and detect suspicious patterns. Following, they sort the resulted cosine distance data for identifying abnormal consumption behavior based on a threshold. 

Moreover, in \cite{Tran2016}, various methods are proposed to resolve the distance-based outlier detection in data streams (DODDS) issue and their performance is compared when detecting anomalies without having any distributional assumptions on power consumption observations. In a similar way, in \cite{Huo2019}, Huo et al. develop an distance-based abnormality detection method, in which a time-space trade-off strategy has been deployed for reducing the computational cost.
While in \cite{Fan2009}, a resolution-based outlier factor (ROF) method is proposed to detect anomalies in large-scale datasets. It mainly focuses on analyzing the distances of both local and global features to effectively detect anomalous data. In \cite{Mao8602251}, the energy consumption anomaly detection process is performed using an isolated forest (iForest) model. The latter has been proposed by Liu et al. as a competitive method to ROF and local outlier factor (LOF) algorithms \cite{Liu4781136,10.1145/2133360.2133363}.

\vskip2mm

\noindent \textbf{F2. Time-series analysis:} because power consumption data are considered time-series footprints, it is logical that many studies have focused on formulating the anomaly detection issue such as to find anomalous observations based on standard signal analysis \cite{8291600Ouyang}. Specifically, this kind of anomaly detection relies on detecting unexpected spikes, level shifts, drops and irregular signal forms. For example, in \cite{Lee2019}, seasonal trend decomposition using locally estimated scatterplot smoothing (LOESS) is proposed to detect anomalous consumption points, in which a seasonal-trend decomposition scheme based on LOESS is introduced. It helps in splitting the power consumption time series samples into three components defined as seasonal, trend, and residue \cite{gao2020robusttad}.

On the other side, it is worth noting that most of the anomaly detection schemes pertaining to this class are based on short-term time-series (STTS) analysis. In this line, a log analysis of power consumption time-series patterns is conducted in \cite{Qiu8377577} to detect real-time anomalies in early warning systems. 
Similarly, \cite{Petladwala8683671}, a feature extraction based abnormality detection scheme is proposed using canonical correlation. It can help in detecting the anomalies in different kinds of buildings, such as households, work spaces and industrial zones. In \cite{Andrysiak2017}, abnormalities occurring in smart meters data are identified using time-series analysis, in which Cook's distance is deployed over a thresholding process to decide whether an observation is normal or abnormal.
In the same vein, in \cite{Ouyang2017}, a hierarchical feature extraction method is proposed in order to capture energy consumption anomalies in time-series consumption data due to electricity stealing. While in \cite{Zyabkina8440460}, to identify the abnormal consumption behavior, the authors analyze different STTS features that could offer valuable details about deviations from a typical behavior.

On the flip side, other techniques use rule-based algorithms to analyze time-series data and detect anomalous power consumption \cite{FLAIRS1715443,Lipcak8859779}. For example, in \cite{WANYEN20191}, Yen et al. introduce a rule-based approach to analyze the phase voltages and then decide which are the anomalous patterns using an ensemble of rules. In the same direction, in \cite{YIP2018189}, a rule-based algorithm is combined with a linear programming approach to detect anomalous electricity consumption and hence identify the locations of potential energy theft attacks and/or faulty meters. In \cite{PENA2016242,Jain8918136}, the detection of anomalous power consumption is performed using a rule-based algorithm, which is elaborated based on machine learning methods and the knowledge of energy saving experts. An ensemble of energy saving parameters is then introduced to track abnormalities. While in \cite{Linda6309297}, a rule-based algorithm is combined with an improved nearest neighbor clustering approach to identify potential abnormal power consumption behaviors.
In \cite{HimeurCOGN2020}, a micro-moment based algorithm is proposed to detect two kinds of power consumption anomalies, which are due to (i) excessive power consumption, and (ii) consumption while the end-users are outside. The latter is responsible of wasting a large amount of energy for a set of appliances, such as the air conditioner, heating system, fan, light lamp and desktop/laptop.

\vskip2mm

\noindent \textbf{F3. Density-based:} refers to anomaly detection methods that investigate the density of each power consumption pattern and those of its neighborhood. Moving forward, a power observation is considered as anomalous if it has a lower density compared to its neighbors \cite{Chen2011AAAI}. Various techniques have been proposed in this regard; among them LOF that attempts to derive a peripheral observation by using density of its surrounding space \cite{10.1145/335191.335388}; cluster-based local outlier factor (CBLOF) that relies on detecting the anomalies using the size of its power consumption clusters, and the density between each power observation and its closest cluster \cite{HE20031641}; local density cluster-based outlier factor (LDCOF) that represents an improved version of CBLOF, in which it applies a local density concept when allocating anomaly scores\cite{giannoni2018anomaly}. In this context, in \cite{Zhou8273883}, a density-based spatial clustering of applications with noise (DBSCAN) approach is introduced to detect anomalous power consumption in a wind farm environment. Overall, density-based anomaly detection has been widely investigated in other fields, such as activity monitoring \cite{kim2013application}, machine fault detection \cite{hou2017data}, financial and banking systems \cite{ahmed2016survey}, etc., their application to detect abnormal energy usage has not been very successful since other kinds of anomalies exist. Specifically, density based schemes could only identify energy consumption outliers based on analyzing energy consumption levels without the possibility to detect other abnormalities, e.g. energy consumption of some appliances (e.g. television, air conditioner, lamp, fan, etc.) while the end-user is absent. 

\vskip2mm

\noindent \textbf{F4. Graph-based:} before applying graph-based methods to detect power consumption abnormalities, consumption data should be converted into a graph-based structure. Because there are no common standards to model this kind of data, researchers use various schemes to design such a representation. For instance, the authors in \cite{Mookiah2017,Akoglu2015}, consider the house, power generator, electric network, rooms, and appliances as nodes; and edges stand for the existing connection between a specific room and the operation of an appliance. Following, abnormalities resulting in a structural change of the graph topology are detected, while a graph-based abnormality is defined as an unforeseen deviation to a normative pattern.

Different graph-based abnormality detection (GBAD) algorithms have been proposed \cite{10.1145/2063576.2063749}, where abnormal observations of structural data are identified in the information representing entities, actions and relationships. In \cite{RAHMANI201489}, the authors propose a graph-based method to discover contextual anomalies in sequential data. Explicitly, the nodes of the graph are clustered into different categories, where each class includes only similar nodes. Following, anomalies are detected via checking if  adjacent observations pertain to the same class or not. Similarly, in \cite{FARAG2019}, a parallel graph-based outlier detection (PGBOD) technique is introduced for identifying power abnormalities, in which data are processed in parallel before extracting abnormal patterns. 

\subsubsection{Hybrid learning (H)}
Annotating normal power consumption is much easier than labeling anomalous patterns, consequently, hybrid or semi-supervised anomaly detection has been adopted in several frameworks \cite{ruff2019deep}. It leverages available annotated normal footprints (having labels) and pertaining to the positive class to identify abnormalities from the negative class. This is the case of deep autoencoder (DAE) architecture when it is only applied to learn normal consumption patterns (with no anomalies). Accordingly, using enough training consumption observations from the normal category, the autoencoder could generate low reconstruction errors for normal observations over abnormal patterns \cite{FAN20181123}. 

In \cite{Wang8848388}, a semi-supervised support vector machine (semi-SVM) based anomaly detection solution is proposed, where a small number of annotated power consumption patterns are required to train the learning model. This system can also generate alarms if suspicious consumption patterns are detected, which are different to usual energy consumption habits of the end-users. While in \cite{Song2017}, DAE and ensemble k-nearest neighbor graphs (KNNG) are combined to develop a semi-supervised anomaly detection system, in which only normal events with their labels are used to train the learning model.

\subsubsection{Other techniques (O)}
In addition to what has been presented in the aforementioned subsections, there are other types of anomaly detection techniques that are built on completely different strategies, including visualization and compressive sensing.

\vskip2mm

\noindent \textbf{O1. Visualization:} offers effective tools to comprehend consumption behavior of end-users through mapping consumption footprints with visual spaces. In this line, visual experts make use of perceptual skills for helping end-users perceive and decipher their consumption patterns within data. Moreover, visualization of load usage footprints could efficiently aid in detecting anomalous consumption behaviors, faulty appliances and suspicious consumption fingerprints that may be due to energy theft attacks. Accordingly, this enables end-users and energy managers to fix related issues and reduce wasted energy. 

For example, in \cite{JANETZKO201427}, the authors propose an anomaly detection framework based on providing various time series visualization schemes, which helps in analyzing and understanding energy consumption behavior. Moreover, it also enables visualizing resulting anomaly scores to direct the end-user/analyst to important anomalous periods.
In the same way, an interactive visualization approach that helps in capturing power consumption anomalies is proposed in \cite{Cao8022952}. It focuses on analyzing and visualizing spatio-temporal consumption footprints gleaned using various streaming data sources. This method has been developed with respect to two prerequisites of real-world anomaly detection systems, which are the online monitoring and interactivity.
Moreover, an interactive dashboard is designed in \cite{CHOU2017711} using an early warning application, which can automatically analyze energy consumption footprints and provide end-users with timely abnormal consumption visualizations based on data recorded from smart meters and sensors. While in \cite{CAPOZZOLI2018336}, a graphical visualization tool for supporting the detection and diagnosis of power consumption abnormalities using a rule-based approach is proposed.  

\vskip2mm

\noindent \textbf{O2. Compressive sensing:} represents a signal processing strategy for effectively analyzing and reconstructing time-series data using their sparsity. It has been widely used in different research fields, such as facial recognition, holography and monitoring of bio-signals. In addition, compressive sensing puts all the appropriate qualities to detect anomalies in energy consumption \cite{Saragadam7951482}. For instance, in \cite{Xia6089743}, the authors prove the relevance of applying compressive sensing in sparse anomaly detection, it relies on the fact that the number of anomalous patterns is generally smaller than the total number of events. In the same direction, in \cite{Wang2016}, separable compression sensing is combined with PCA to identify anomalous power data. In \cite{Levorato6250561}, anomalous events in smart grid are detected using a sparse approximation paradigm.

\subsection{Anomaly detection level}
The anomaly detection level of power consumption data plays a major role in developing effective solutions because it describes either the level of resolution in which power anomalies have been detected and treated. Correspondingly, tailored recommendations could be generated to resolve the associated issues and promote energy efficient behavior. 

\vskip2mm

\noindent{\textbf{L1. Aggregated level:}} it refers to detecting anomalous power consumption using data of the main supply in a specific building, i.e. without any information about individual consumption of the different appliances connected to the electrical network. Although this kind of anomaly detection has been used in various works, it has the main drawback of not being able to provide the end-user with information about which appliance is responsible for a specific anomaly.

\vskip2mm

\noindent{\textbf{L2. Appliance level:}} it stands for the case where anomaly detection is performed using appliance power consumption data gathered using individual sub-meters. This kind of anomaly detection is widely adopted because it supports a fine-grained tracking of abnormalities occurring during the operation of each electrical device \cite{Rashid8683792}.

\vskip2mm

\noindent \textbf{L3. Spatio-temporal level:} much attention has been devoted recently to the collection of continuous spatio-temporal power consumption patterns from different devices and sources. This affords new opportunities to timely understand consumption fingerprints in their spatio-temporal context \cite{Liu2017taylor,Munawar7926701}. Overall, detecting anomalous consumption behaviors using conventional data collection methods present considerable challenges since the boundary between normal and anomalous observations is not obvious. Therefore, a straightforward solution to those challenges is to interpret consumption abnormalities in their multifaceted and spatio-temporal context. Specifically, detecting abnormal consumption related to specific hours in the day, or what are the severe days presenting anomalous consumption and how to identify them in the timestamps (weekdays, weekends, holidays, etc.) will be valuable to provide end-users with a personalized feedback to reduce wasted energy \cite{Yang2018bigdata,BOSMAN201741}.

\subsection{Applications}

The applications of anomaly detection of energy consumption in buildings are no longer limited to energy efficiency, but they are finding themselves in various novel application contexts. Explicitly, they could be used for detecting (i) abnormal consumption behaviors, (ii) faulty appliances, (iii) occupancy information, (iv) non-technical losses, and (v) at-home elderly monitoring. In addition, the same anomaly detection system, within a building can be used for multiple applications without the need for installing other systems (e.g. to detect occupancy or non-technical losses). Therefore, this could effectively reduce the hardware implementation costs and decrease the complexity of installed systems.


\vskip2mm

\noindent \textbf{A1. Detection of abnormal behavior of end-users:} it is the main application for which anomaly detection has been proposed since the final objective is to reduce wasted energy and promote sustainable and energy efficiency behaviors \cite{HimeurCOGN2020,PENA2016242}. In this context, detecting anomalous consumption behavior of end-users allows a better and accurate assessment of power usage, which can be translated into providing them with useful and personalized recommendations \cite{Enetics2020,Dilraj8929235}.

\vskip2mm

\noindent \textbf{A2. Detection of faulty appliance:} using various kinds of appliances at indoor environments has made people's lives more convenient. However, these electrical appliances could be faulty in different ways or could suffer from inefficiencies, and hence leading to several issues, such as the events resulting in a massive energy waste and triggering electrical fires \cite{10.1145/3276774.3276797,Kabler2014}. To that end, detecting faulty appliances and providing the end-users with customized recommendations to replace them is of significant importance in reducing the operation cost and boosting energy saving in buildings \cite{Rashid8683792,YIP2018189}.

\vskip2mm

\noindent \textbf{A3. Occupancy detection:} detecting whether a building or one of its parts is occupied by the end-users is essential to allow a set of building automation tasks. Although actual tools for detecting the indoor occupancy typically need to install specialized sensors, including passive-infrared sensors (PIR), reed switches actuated by magnets, or cameras, their installation is very costly and further labor charges could be added for maintenance \cite{Sardianos2020iciot,LAAROUSSI2020102420}. Therefore, a solution to overcome the high-cost pitfall is to explore the aptitude of electrical sub-meters, which are installed in most of the houses around the globe to detect occupancy patterns \cite{10.1145/2528282.2528295,Akbar7248381}. For example, the authors in \cite{Gao8614235} investigate both appliance specific and aggregated load usage footprints to detect the occupancy of residents \cite{Violatto8706523}.

\vskip2mm

\noindent \textbf{A4. Non-technical loss detection:} it mainly refers to (i) detecting unintentional sub-meters' dysfunctions and electricity theft attacks attempting to bypass sub-meters; (ii) braking and/or stopping sub-meters; (iii) identifying faulty sub-meters' records; and (iv) capturing appliances having illegal connections \cite{Depuru5772466,Nabil8545748}. Non-technical loss in energy consumption has negatively affected most of the economies over the globe \cite{Yeckle8367753}. For instance, more than 10\% of produced energy could be lost every year in Europe due to non-technical loss and billions of dollars are lost every year because of theft energy attacks \cite{YIP2018189,Krishna7579759}. To that end, detecting non-technical-loss and electricity theft have been introduced as an information technology related challenge, which requires novel methods based on AI, data mining and forecasting \cite{HOCK2020100290,Korba8807832}. Moreover, separating between behavioral consumption anomalies, fraud and unintentional consumption deviations is reported as a current research trend to provide an accurate feedback to end-users and energy providers \cite{Albiero2019,Jain8918136}.

\vskip2mm

\noindent \textbf{A5. At-home elderly monitoring:} modern societies face significant issues with the monitoring of their elderly people at home environments \cite{Visconti8791399}. This problem could have considerable social and economic effects. However, one solution to overcome it is via (i) monitoring appliance consumption of elderly people in real-time; (ii) identifying abnormal consumption behaviors that could be occurring due to some critical situations (e.g. falls); and (iii) predicting faulty operations of some appliances, which can result in dangerous situations (e.g. floods or gas leaks) \cite{Patrono8115547,Patrono8448304}.  

\subsection{Computing platforms}
As presented previously, most of the anomaly detection methods have been built upon the use of machine learning techniques. However, although the use of these approaches has dived the development of anomaly detection technology, it requires serious challenges of computing resources, data processing speed and scalability. In this regard, describing and discussing available solutions used to implement anomaly detection systems is essential to understand the current challenges.

\begin{itemize}

\item \textbf{P1. Edge computing platforms:} refer to distributed computational models that allow to drop the computing resources and information storage capabilities close to the end-user application, where it can directly be used, e.g. in energy consumption applications this can be done on the smart sensor platforms or smart plug devices, as it is the case in (EM)$^{3}$ \cite{himeur2020emergence}. Specifically, a smart plug is being developed to incorporate different sensors to collect consumption and contextual data along with a micro-controller to pre-process data, segregate the main consumption signal into device specific footprints, and detect abnormal behaviors. This helps in improving output, accelerating data processing and saving bandwidth \cite{Zhang9179818}.

\item \textbf{P2. Fog computing platforms:} stand for decentralized computational infrastructures, where data pre-processing, computing, storage and analysis are conducted in the layer located between the data collection devices and the cloud \cite{s20010122}. In this line, the computational ability of the anomaly detection solution is carried out close to both the data recording devices and the cloud, in which data are produced and handled \cite{app9194193}.

\item \textbf{P3. Cloud computing platforms:} concern the cases when the computing and storage resources are ensured using distant servers, in which the end-users deploying the anomaly detection solutions are required to connect them through an internet link to be able to execute the anomaly detection algorithms \cite{Liu2016lambda}. Put differently, the platforms used to implement these algorithms become as the access points for running the anomaly detection applications and visualize the data held by the servers. The cloud architectures are described by their flexibility, which allows the providers to constantly adjust the storage capability and computing power to the end-users' requirements \cite{LIU2020107212}.

\item \textbf{P4. Hybrid computing platforms:} refer to the cases where the computing power is guaranteed by various layers, including the cloud, fog and edge as explained in \cite{Zhang2017}. In this context, based on the computing requirement of the anomaly detection solution and the existing computing resources, the algorithms could be executed either on the edge and/or fog when they need a low computation cost, otherwise they could be implemented in the cloud when high computing cost is required \cite{Anjomshoaa8361419,Izumi8116677}.

\end{itemize}

Table \ref{SummaryAD}, presents a comparison of several aforementioned anomaly detection frameworks in building energy consumption. They are compared with reference to various  parameters, such as the (i) application scenario, (ii) category, (iii) implemented technique, (iv) learning process, (v) computing platform used (or required) to implement the anomaly detection algorithm, (vi) privacy preservation, and (vii) sampling rate. This helps in easily understanding the properties of each framework and difference between existing solutions.

\subsection{Example of anomaly detection using AI}
In order to explain how anomalies of energy consumption have been considered in the literature and how AI could be used to detect abnormal usage, we present in this section, three different scenarios for anomaly detection using (i) AI-based prediction, (ii) AI classification of energy micro-moments and occupancy data and (iii) one-class classification of energy data. It is worth noting that with the use of AI, it becomes possible to detect more advanced kinds of anomalies using other types of data, such as occupancy patterns and ambient conditions. 

\vskip3mm

\begin{figure*}[!t]
\centering
\includegraphics[width=\columnwidth]{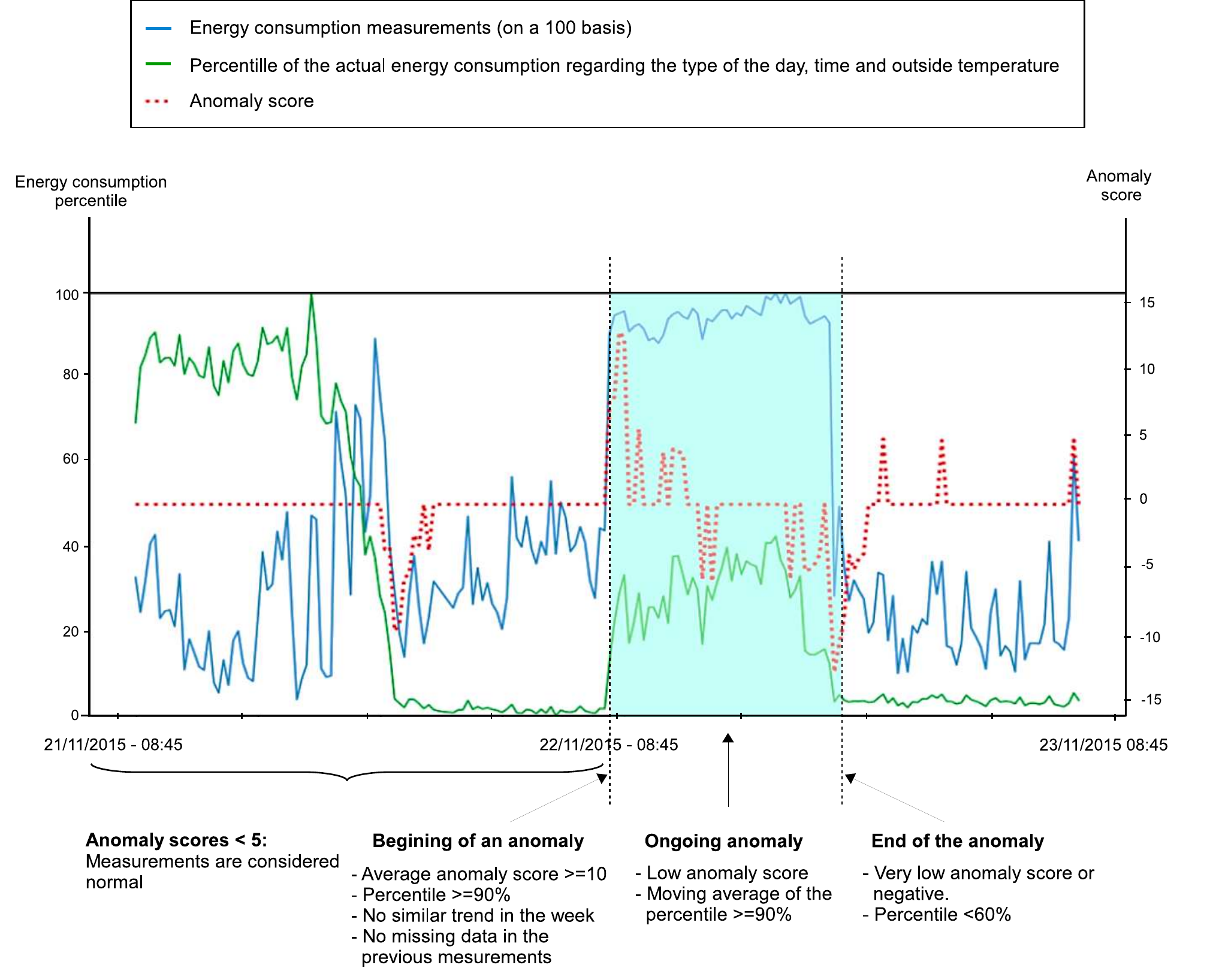}
\caption{Example of anomaly detection in time-series energy consumption using AI-based prediction applied on anomaly detection dataset provided in \cite{drivendata2020}.}
\label{time-series} 
\end{figure*}

\textbf{Scenario I. Anomaly detection using AI-based prediction:}\\ 
In \cite{drivendata2020}, a power consumption dataset is provided to validate anomaly detection algorithms. AI tools for predicting future energy consumption are combined with a rule-based algorithm for detecting anomalous energy consumption. This method relies on using an RNN model to predict energy consumption for the next timestamp and then the difference between real and predicted consumption is calculated to measure the level of \enquote{surprise}. In this regard, if a significant gap is detected, either (i) an anomalous energy usage behavior has occurred, or (ii) the model has made a mistake. Moving forward, the investigation is then continued using a rule-based algorithm and by filtering the abnormalities identified by the predictive system with reference to a set of statistical criteria. The rule-based algorithm helps in detecting percentile of power consumption for each timestamp with regard to the (i) hour, (ii) outside	temperature,	and	(iii) type of day (working day vs. holiday). Fig. \ref{time-series} illustrates an example of anomalies detected when analyzing time-series energy consumption. 

\vskip3mm

\begin{figure}[!t]
\centering
\includegraphics[width=0.875\columnwidth]{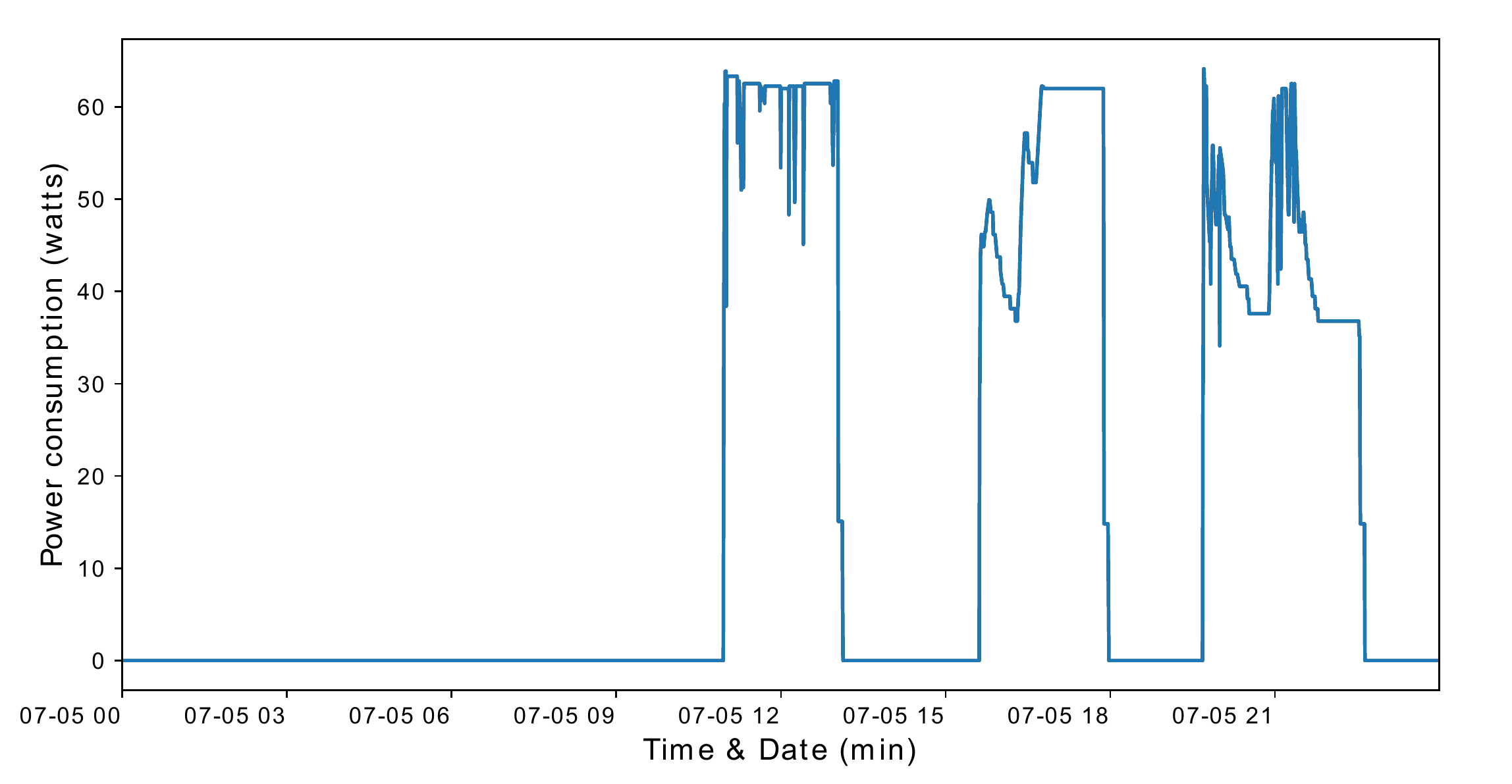}
\includegraphics[width=0.875\columnwidth]{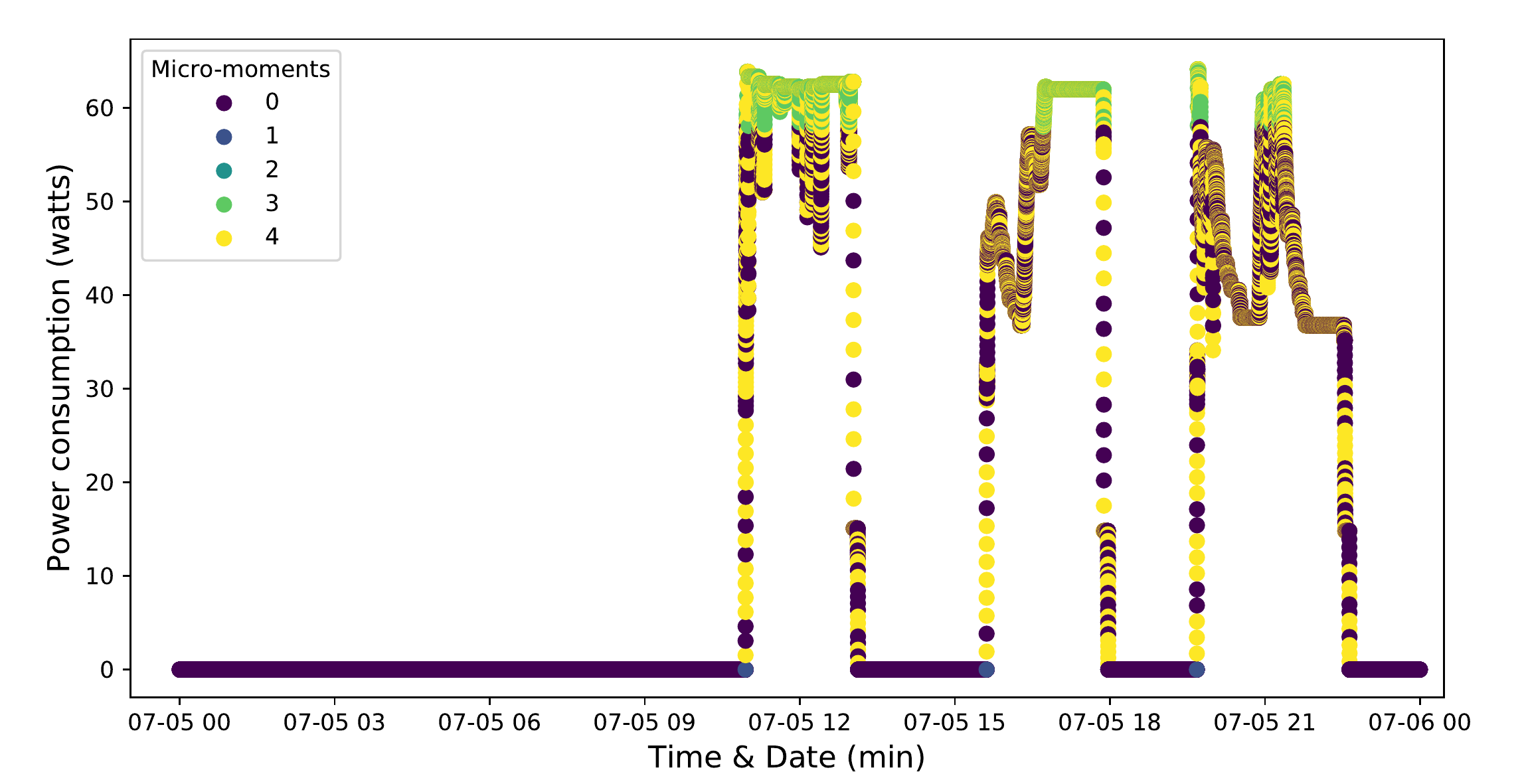}
\caption{Example of anomaly detection based on DNN and micro-moment analysis with reference to energy data and occupancy patterns. These visualization plots are derived using DRED dataset (for the case of a television): time-series energy consumption traces, and bottom) micro-moment detection scheme based on deep learning \cite{Himeur2020IJIS-AD}.}
\label{VisPlot} 
\end{figure}

\textbf{Scenario II. Anomaly detection using AI and micro-moment analysis:} \\
In the (EM)$^3$ framework \cite{Himeur2020IJIS-AD}, anomalous energy consumption is detected by analyzing energy consumption footprints and occupancy patterns using micro-moments analysis. Following, a DNN model is deployed to automatically classify each consumption observation as normal or abnormal. Specifically, energy consumption samples are clustered into five classes; three of them are named \enquote{class 0: good usage}, \enquote{class 1: turn on a device} and \enquote{class 2: turn off a device}, which represent normal usage; and the other two classes are called \enquote{class 3: excessive consumption} and \enquote{class 4: consumption while outside}, which refer to abnormal usage. Fig. \ref{VisPlot} illustrates an example of time-series energy traces collected in the DRED dataset for the case of a television \cite{DRED2015}, and the corresponding normal and abnormal energy patterns identified using a DNN model and micro-moment analysis. Because of occupancy data consideration in the anomaly detection stage, it was possible to detect a new consumption anomaly that corresponds to the absence of the end-user when the television is on (this abnormality could also be considered for other specific appliances, such as the air conditioner, heater, fan, etc.), this was not possible using conventional anomaly detection techniques that are based only on analyzing energy fingerprints.

\begin{landscape}
\begin{table} [t!]
\caption{Summary of research frameworks conducted in energy consumption anomaly detection.}
\label{SummaryAD}
\begin{center}

\begin{tabular}{llllllll}
\hline
{\small Reference (year)} & {\small Application} & {\small Category} & 
{\small Implemented technique} & {\small Learning } & {\small Computing} & 
{\small Privacy} & {\small Sampling} \\ 
&  &  &  & {\small process} & {\small platform} & {\small preservation} & 
{\small rate} \\ \hline
{\small \cite{Yeckle8367753} (2017)} & {\small A1} & {\small U1} & {\small %
MNN and k-means clustering} & {\small Unsupervised} & {\small -} & {\small -}
& {\small 1 hour} \\ 
{\small \cite{Ghori2020} (2020)} & {\small A1} & {\small U2} & {\small OCRF}
& {\small Unsupervised} & {\small P1} & {\small -} & {\small 1 hour, } \\ 
{\small \cite{Himeur2020IJIS-AD} (2020)} & {\small A1} & {\small U2,U3} & 
{\small OCSVM, DBSCAN, LOF, LDA, IKNN} & {\small Supervised} & {\small %
P1,P3,P4} & {\small No} & {\small 1 sec, 3 sec } \\ 
{\small \cite{Weng8574884} (2019)} & {\small A1} & {\small S1} & {\small %
Autoencoder and RNN } & {\small Unsupervised} & {\small P3} & {\small -} & 
{\small 1 min} \\ 
{\small \cite{Pereira8614232} (2018)} & {\small A1} & {\small S1} & {\small %
Variational recurrent autoencoder} & {\small Supervised} & {\small P3} & 
{\small -} & {\small 15 min} \\ 
{\small \cite{10.1155/2019/4136874} (2019)} & {\small A4} & {\small S1} & 
{\small CNN and random forest} & {\small Supervised} & {\small P3} & {\small %
No} & {\small 1 hour} \\ 
{\small \cite{Zheng8233155} (2018)} & {\small A4} & {\small S1} & {\small CNN%
} & {\small Supervised} & {\small P3} & {\small -} & {\small 1 hour, 1 day}
\\ 
{\small \cite{en13010130} (2020)} & {\small A1} & {\small S1} & {\small %
Recurrent GAN} & {\small Supervised} & {\small P3} & {\small No} & {\small 1
hour} \\ 
{\small \cite{daSilva8702152} (2019)} & {\small A1} & {\small S1} & {\small %
RNN and negative selection} & {\small Supervised} & {\small P2} & {\small No}
& {\small 30 min} \\ 
{\small \cite{Chahla2019} (2019)} & {\small A1} & {\small S1} & {\small RNN
and and K-means} & {\small Unsupervised} & {\small P3} & {\small -} & 
{\small 1 hour} \\ 
{\small \cite{XU2020} (2020)} & {\small A1} & {\small S1} & {\small RNN and
quantile regression} & {\small Unsupervised} & {\small P4} & {\small -} & 
{\small 1 hour} \\ 
{\small \cite{WANG2020114145} (2020)} & {\small A1} & {\small S1} & {\small %
ANNs and ARIMA} & {\small Supervised} & {\small P1} & {\small No} & {\small %
1 hour} \\ 
{\small \cite{Mulongo2020} (2020)} & {\small A2} & {\small S1} & {\small MLP}
& {\small Supervised} & {\small P1,P2} & {\small -} & {\small 1 hour} \\ 
{\small \cite{Gaur8709671} (2019)} & {\small A1} & {\small S2} & {\small %
Linear regression \ + rule-based algorithm} & {\small Supervised} & {\small %
P1} & {\small No} & {\small 1 hour} \\ 
{\small \cite{info8040151} (2017)} & {\small A1} & {\small S2} & {\small %
autoregressive prediction} & {\small Semi-supervised} & {\small P1,P2} & 
{\small -} & {\small 30 min} \\ 
{\small \cite{HOCK2020100290} (2020)} & {\small A4} & {\small S3} & {\small %
Bayes algorithms } & {\small Supervised} & {\small P1} & {\small No} & 
{\small 5 min} \\ 
{\small \cite{Saqaeeyan2020} (2020)} & {\small A4} & {\small S3} & {\small %
Bayesian networks} & {\small Supervised} & {\small P1} & {\small No} & 
{\small 15 min} \\ 
{\small \cite{Liu2016lambda} (2016)} & {\small A4} & {\small S3} & {\small %
Gaussian distribution } & {\small Supervised} & {\small P1} & {\small -} & 
{\small 1 hour} \\ 
{\small \cite{CAPOZZOLI2018336} (2018)} & {\small A1} & {\small O2} & 
{\small Graphical visualizatiuon } & {\small Unsupervised} & {\small P3} & 
{\small -} & {\small 30 min, 1 hour} \\ 
{\small \cite{Korba8807832} (2019)} & {\small A4} & {\small S4} & {\small SVM%
} & {\small Supervised} & {\small P1,P2} & {\small No} & {\small 1 hour} \\ 
{\small \cite{HimeurCOGN2020} (2020)} & {\small A1} & {\small S4,S1} & 
{\small SVM, KNN, decision tree, EBT, DNN} & {\small Supervised} & {\small %
P1,P3,P4} & {\small No} & {\small 1/6 sec, 1 sec} \\ 
{\small \cite{TOUZANI20181533} (2018)} & {\small A1} & {\small E1} & {\small %
GBM} & {\small Supervised} & {\small P1} & {\small -} & {\small 15 min} \\ 
{\small \cite{Albiero2019} (2019)} & {\small A4} & {\small E1} & {\small GBM
and grid search} & {\small Supervised} & {\small P1} & {\small -} & {\small -%
} \\ 
{\small \cite{ARAYA2017191} (2017)} & {\small A1} & {\small E2} & {\small %
Bootstrap aggregation} & {\small Supervised} & {\small P1,P2} & {\small -} & 
{\small 5 min} \\ 
{\small \cite{Wun8974857} (2019)} & {\small A1} & {\small E2} & {\small %
SCiForest} & {\small Supervised} & {\small P1} & {\small No} & {\small 30 min%
} \\ 
{\small \cite{Yijia7581646} (2016)} & {\small A1} & {\small F1} & {\small %
Distance-based approach} & {\small Unsupervised} & {\small P1,P2} & {\small -%
} & {\small -} \\ 
{\small \cite{Petladwala8683671} (2019)} & {\small A1} & {\small F2} & 
{\small Time-series analysis} & {\small Supervised} & {\small P1,P2} & 
{\small No} & {\small 1 min} \\ 
{\small \cite{FAN20181123} (2018)} & {\small A1} & {\small H} & {\small DAE}
& {\small Semi-supervised} & {\small P3,P4} & {\small -} & {\small 30 min}
\\ 
{\small \cite{Wang8848388} (2019)} & {\small A1} & {\small H} & {\small %
Semi-SVM} & {\small Semi-supervised} & {\small P1} & {\small -} & {\small 1
hour} \\ 
{\small \cite{Patrono8115547} (2017)} & {\small A5} & {\small F1} & {\small %
Rule-based algorithm} & {\small Unsupervised} & {\small P1} & {\small No} & 
{\small 30 sec} \\ 
{\small \cite{Violatto8706523} (2019)} & {\small A4} & {\small F2, S1} & 
{\small Time-frequency features + OCRF} & {\small Supervised} & {\small P1,P2%
} & {\small -} & {\small 10 min, 1 hour} \\ 
{\small \cite{10.1145/3276774.3276797} (2019)} & {\small A2} & {\small S3} & 
{\small Rule based statistical model } & {\small Unsupervised} & {\small P1}
& {\small No} & {\small 10 min} \\ 
{\small \cite{Nabil8746794} \ (2019)} & {\small A4} & {\small S1} & {\small %
CNN} & {\small Supervised} & {\small P3} & {\small Yes} & {\small 30 min} \\ 
{\small \cite{Yao8661537} (2019)} & {\small A4} & {\small S1} & {\small CNN}
& {\small Supervised} & {\small P3} & {\small Yes} & {\small -} \\ \hline
\end{tabular}

\end{center}
\end{table}
\end{landscape}

\vskip3mm

\begin{figure}[!t]
\centering
\includegraphics[width=0.75\columnwidth]{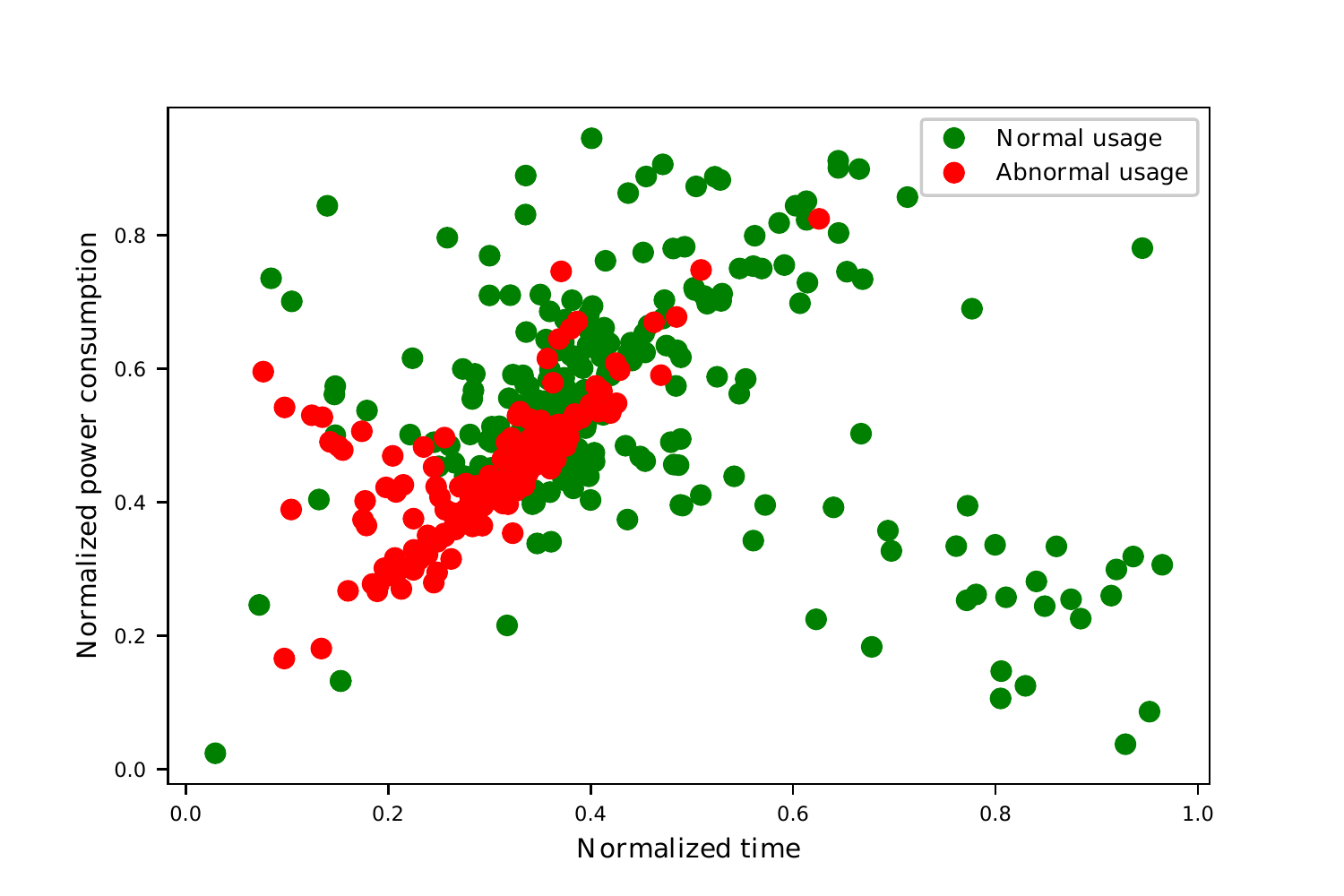}
\caption{Example of anomaly detection in time-series energy consumption using autoencoders applied on DRED dataset \cite{DRED2015}.}
\label{auto-encoder} 
\end{figure}

\textbf{Scenario III. Anomaly detection using one-class classification:}\\
Another important anomaly detection solution is based on conventional one-class classification, which has been widely utilized in other applications. Fig. \ref{auto-encoder} shows an example of an anomaly detection of energy consumption applied on DRED dataset \cite{DRED2015} using the one-class autoencoder. It has clearly been seen that this scheme divides energy observation into two main classes based on the analysis of the energy consumption levels in a new representation space, in which the power and time have been normalized \cite{himeur2020detection}.

\section{Critical analysis and discussion} \label{sec3}
\subsection{Discussion}
Anomaly detection in building energy consumption is of paramount importance to developing powerful energy management systems, identifying energy theft attacks, inefficiencies and negligence. However, in most cases it is difficult to separate consumption abnormalities from the normal usage deviations occurring owing to seasonal changes and variation of personal settings (e.g. holidays, family parties, unexpected changes of due new circumstances, etc.). Moreover, one of the limitations of available anomaly detection methods is related to the fact that diverse unidentified context data, including seasonal changes, could impact the power usage of end-users in a manner to be as abnormal when existing time-series based anomaly detection techniques are used. In addition, a set of important findings can summarized as follows:

\begin{itemize}

\item AI-based solutions focus mainly on developing real-time or near real-time (e.g. at a hourly sampling rate or lower) although they can also provide an insight analysis for time long periods (e.g. days, weeks, months and years). This is due to the capability of AI to analyze big data, especially when high frequency sampling rates are considered and also thanks to the IoT devices, smart-meters and smart sensors, which help tremendously in collecting accurate data. On the other hand, this represents the main difference between actual AI-based anomaly detection techniques and those used twenty or thirty years ago, where it was not possible to process data in real-time or near real-time. In addition, almost all the reviewed frameworks have focused on the analysis of power consumption data either on kWh or Wh. This depends on whether the anomaly detection has been conducted at the aggregated-level (using kWh) or an appliance-level (using Wh).

\item Most of existing approaches of anomaly detection in energy consumption attempt only to flag out power samples that are remarkably higher or lower than usual consumption footprints, as it is the case in other applications, such as bank card fraud detection, network intrusion detection and electrocardiogram anomaly detection. Unfortunately, this is not the correct case to detect anomalous power consumption because the definition of anomaly in energy consumption can be quite different, other kinds of anomalies are available and their detection requires other information sources, e.g. occupancy patterns, and appliance operation data.

\item By using AI, it becomes possible to develop real-time or near real-time energy consumption anomaly detection systems, which could identify timely anomalous usage and alert the end-users by sending warnings and notifications. Accordingly, recommender systems could then be deployed to help the end-users with a better decision-making to reduce their wasted energy through providing them with personalized and contextual recommendations. For instance, the EM$^{3}$ project \footnote{\url{http://em3.qu.edu.qa/}} combines anomaly detection and a recommender system to help end-users in reducing their wasted energy using both real-time or near real-time strategies.

\item According to recent works \cite{Rashid8683792,RASHID2019796}, using aggregated-level consumption data is not the best way to detect anomalies of energy consumption because they are general and can not provide precise information on the causes of each anomaly. Therefore, using appliance-level data generated either by sub-meters or using NILM systems is more appropriate since this helps in detecting the anomalies of each appliance \cite{Gaur8709671,HIMEUR2020115872}.

\item In some cases, the entirety of a given power consumption behavior could be considered as abnormal and not only some specific observations, which make it difficult to detect the exact anomalous parts. Therefore, this requires comparing current consumption footprints with the past and ideal consumption cycles and not only using outlier detection algorithms, which can detect the anomalies at the sample level.

\item In terms of the effectiveness of existing methods, although unsupervised anomaly detection is easy to implement since it does not require annotated datasets to learn the anomalies, it presents serious drawbacks because it can only detect one kind of anomalies, which is related to excessive consumption. This is also the same with ensemble methods and feature extraction-based techniques. In contrast, supervised methods are not very popular as unsupervised ones as they require using labeled datasets to learn the abnormalities. However, using methods pertaining to this category allows to detect other types of anomalies. This is because they could be defined a priori by human experts using training data collected from different sources, e.g. consumption footprints, occupancy patterns, indoor conditions and appliance operation parameters.

\item In terms of the computing resources, most of the deep learning based anomaly detection frameworks require high-performance computing capabilities to conduct the learning process. Therefore, most of them use cloud computing to integrate and manage large datasets. While for conventional machine learning based anomaly detection, edge and fog computing have been successfully used in various frameworks and applications.

\item Privacy preservation: developing anomaly detection systems to promote energy saving in buildings is of paramount importance at all levels of the society. This can be performed using local and temporal fine-grained records of power consumption fingerprints, occupancy patterns and ambient conditions to identify abnormalities and unnecessary power consumption \cite{Salinas7087399}. Unfortunately, using this kind of fine-grained records enables disclosing information on the presence of the end-users based on their energy usage footprints. In this context, we have noticed that the privacy preservation has been ignored or not reported in most of the anomaly detection frameworks, only very few have tried to touch on this issue \cite{Nabil8746794,Yao8661537}.

\end{itemize}

\subsection{Relevance of AI-based anomaly detection techniques}
The relevance and robustness of AI-based anomaly detection does not rely only on the accuracy of detecting anomalous energy usage, but also on the type and number of the consumption abnormalities that could be detected. In this regard, it was clear that most of the unsupervised anomaly detection techniques (i.e. clustering, one-class classification and dimensionality reduction) could detect only one type of energy usage anomalies, which corresponds to excessive energy consumption. This is because they are based on identifying rare consumption observations or outliers, which raise suspicions by differing considerably from the majority of the consumption footprints. In addition, they only analyze energy consumption data without considering other relevant factors that impact energy usage, such as occupancy, ambient conditions and users' preferences. On the other hand, supervised anomaly detection presents more advantages since they can be utilized to detect different kinds of energy consumption abnormalities by considering the impact of the presence/absence of end-users, ambient conditions, outdoor weather data and users' preferences on energy usage \cite{Gaur8709671,himeur2020detection}. This was possible through the use of rule-based algorithms to define abnormal consumption and annotate multi-modal datasets. In this context, deep anomaly detection techniques that are based on adopting deep learning models presents promising performance and in terms of the accuracy of detecting abnormal usage and also because of their capability to process and analyze multi-modal data, as described in \cite{HimeurCOGN2020}. Table \ref{Adv-Disadv} presents a summary of relevant AI-based anomaly detection techniques, including their strengths and weaknesses.

\begin{table} [!t]
\caption{A summary of relevant AI-based anomaly detection techniques, including their strengths and weaknesses.}
\label{Adv-Disadv}
\begin{center}

\begin{tabular}{llll}
\hline
{\small Ref.} & {\small Implemented technique} & {\small Advantages} & 
{\small Drawbacks} \\ \hline
{\small \cite{Yeckle8367753} } & {\small MNN and k-means } & {\small No need
for annotated data} & {\small Low detection accuracy, detection of only } \\ 
& {\small clustering} &  & {\small excessive consumption} \\ 
{\small \cite{Ghori2020} } & {\small OCRF} & {\small No need for annotated
data} & {\small Low detection accuracy, detection of only } \\ 
&  &  & {\small excessive consumption} \\ 
{\small \cite{Himeur2020IJIS-AD}} & {\small OCSVM, DBSCAN, LOF} & {\small %
Detection of Two kinds of anomalies, } & {\small Absence of annotated data.}
\\ 
& {\small LOF, LDA, IKNN} & {\small high detection accuracy} &  \\ 
{\small \cite{Weng8574884}} & {\small Autoencoder and RNN } & {\small No
need for annotated data} & {\small Detection of only excessive consumption, }
\\ 
&  &  & {\small \ low detection performance} \\ 
{\small \cite{Pereira8614232}} & {\small Variational recurrent } & {\small %
No need for annotated data} & {\small Difficulty to assess the performance}
\\ 
& {\small autoencoder} &  &  \\ 
{\small \cite{10.1155/2019/4136874} } & {\small CNN and random forest} & 
{\small High anomaly detection  } & {\small Analyze only energy consumption
data, } \\ 
&  & {\small performance} & {\small high computational cost} \\ 
{\small \cite{Zheng8233155}} & {\small CNN} & {\small Capture the abnormal
electricity  } & {\small Lack of annotated data, detection of excessive }
\\ 
&  & {\small usage with a high accuracy} & {\small consumption, high
computational cost} \\ 
{\small \cite{daSilva8702152}} & {\small RNN and negative } & {\small %
Predict excessive consumption  } & {\small Low detection performance, only
excessive } \\ 
& {\small selection} & {\small anomalies} & {\small consumption is detected}
\\ 
{\small \cite{XU2020} } & {\small RNN and quantile } & {\small High
detection performance, no need } & {\small Detection of only excessive
consumption, } \\ 
& {\small regression} & {\small for annotated data} & {\small weak
interpretability} \\ 
{\small \cite{WANG2020114145} } & {\small ANNs and ARIMA} & {\small Anomaly
detection and energy usage } & {\small High training cost, detection of only 
} \\ 
&  & {\small prediction, high detection accuracy} & {\small excessive
consumption} \\ 
{\small \cite{Gaur8709671} } & {\small LR \ + rule-based algo} & {\small Low
training cost} & {\small Difficulty to annotate data, low detection accuracy}
\\ 
{\small \cite{info8040151} } & {\small Autoregressive prediction} & {\small %
Anomaly detection and power usage } & {\small Low prediction performance,
detection of only } \\ 
&  & {\small prediction} & {\small excessive consumption} \\ 
{\small \cite{Liu2016lambda} } & {\small Gaussian distribution } & {\small %
Low training cost} & {\small Low detection performance, lack of annotated }
\\ 
&  &  & {\small data, detection of excessive consumption} \\ 
{\small \cite{HimeurCOGN2020} } & {\small SVM, KNN, DT, } & {\small %
Detection of Two kinds of anomalies, } & {\small Lack of annotated data} \\ 
& {\small EBT, DNN} & {\small high detection performance} &  \\ 
{\small \cite{TOUZANI20181533} } & {\small GBM} & {\small Anomaly detection
and power usage } & {\small Detection of only suspicious consumption } \\ 
&  & {\small prediction} & {\small levels, weak interpretability} \\ 
{\small \cite{Albiero2019} } & {\small GBM and grid search} & {\small Low
training cost} & {\small Low detection performance, weak interpretability,}
\\ 
&  &  & {\small one type of anomalies is detected } \\ 
{\small \cite{ARAYA2017191} } & {\small Bootstrap aggregation} & {\small %
High detection performance} & {\small Difficulty to set the optimal
threshold, detection } \\ 
&  &  & {\small of only suspicious consumption level} \\ 
{\small \cite{Yijia7581646} } & {\small Distance-based approach} & {\small %
Low training cost} & {\small Weak interpretability, low detection
performance } \\ 
&  &  &  \\ 
{\small \cite{Petladwala8683671} } & {\small Time-series analysis} & {\small %
Low training cost} & {\small Low detection accuracy, detection of only } \\ 
&  &  & {\small excessive consumption} \\ 
{\small \cite{FAN20181123} } & {\small DAE} & {\small Anomaly detection and
power usage} & {\small High computation cost, detection of only } \\ 
&  & {\small prediction, high detection accuracy} & {\small excessive
consumption} \\ 
{\small \cite{Wang8848388} } & {\small Semi-SVM} & {\small Anomaly detection
and power usage } & {\small Weak interpretability, detection of only } \\ 
&  & {\small prediction} & {\small suspicious consumption levels} \\ 
{\small \cite{10.1145/3276774.3276797} } & {\small Rule-based statistical  }
& {\small Anomalous appliances detection, } & {\small Low detection
accuracy, detection of only} \\ 
& {\small model} & {\small Low training cost} & {\small excessive consumption%
} \\ 
{\small \cite{Nabil8746794} \ } & {\small CNN} & {\small %
Privacy-preservation, high detection } & {\small Detection of only
suspicious consumption } \\ 
&  & {\small accuracy} & {\small levels, weak interpretability} \\ \hline
\end{tabular} 
\end{center} 

\end{table}


\subsection{Challenges and limitations}
There are several common and domain-specific challenges and limitations of anomaly detection systems in energy consumption, which hinder developing efficient solutions, render their implementation costly and limit their widespread. They can be outlined in the following points:

\begin{itemize}
\item Absence of annotated datasets: among the serious pitfalls to develop and validate abnormality detection schemes is the absence of annotated datasets, which provide labels for both normal and abnormal consumption. Most of the supervised algorithms are validated on a small quantity of data, which can not be considered as comprehensive datasets and are not accessible for the energy research community. Specifically, repositories that label the events of abnormal consumption and their types almost do not exist and its creation is difficult and costly \cite{Gaur8709671}. Therefore, creating various datasets for different kinds of buildings that reflect real consumption behaviors will help effectively the energy research community in testing and improving the detection of consumption abnormalities in different application scenarios \cite{HimeurENB2020}. 

\item Imbalanced dataset: refers to the distribution of anomalies through data classes, i.e. anomalous data might usually be the minority amongst the overall dataset. Indeed, the anomaly data are very rare in reality, forming together with the major normal data an extreme unbalanced set. The class imbalanced characteristic of most of the anomaly detection datasets results in a sub-optimality of the algorithms' performance. Therefore, to deal with this issue, some pre-processing techniques are required, among them (i) using resampling procedures to oversample the minority classes or undersample the majority classes, and (ii) generating synthetic power consumption data \cite{HimeurCOGN2020}. Moreover, in other topics, the anomaly classes are generally represented as minor classes, but in energy consumption this is not always the case, especially if a high energy wasting behavior is observed. In this regards, applying unsupervised anomaly detection methods is less efficient.  

\item Definition of anomalies: traditional definition of an anomaly signifies that an anomalous observation is an outlier or deviant. However, this definition could not be enough to define anomalies in energy consumption because other forms of abnormalities could exist, e.g. keeping an appliance on (i.e. air conditioner, fan, television, etc.) while end-users are outside, keeping windows and doors open when an air conditioner/heating system is switching on, which leads to a high power consumption, etc. Therefore, to efficiently detect anomalies of energy consumption, it is required to analyze not only the power consumption data but also other information sources, including the occupancy patterns, ambient conditions, outside weather footprints and appliance operation parameters. 

\item Sparse labels: on the one hand, the labels denoting whether an instance is normal or anomalous is in many applications time-consuming and prohibitively expensive to obtain. This is especially typical for time series data, where the sampling frequency could reach 1000 Hz or the time could range over decades, generating an enormous amount of data points. On the other hand, anomalous data is often not reproducible and fully concluded in reality.

\item Detecting appliance-level anomalies is still not receiving the necessary attention, although it is more important than detecting aggregated-level anomalies. In effect, a failure in the electronics of an appliance could not only increase energy consumption, but in some cases, other kinds of failures may cause new forms of faulty appliances that could be fatal, e.g. a faulty device can cause an electrical short that sparks a fire.

\item Concept drift: this phenomenon usually occurs in time series data, where the common independent and identically distributed (i.i.d) assumption for machine learning models is often violated due to the varying latent conditions \cite{Liu2014}. Since the observations and relations in power consumption data evolve over time, they should be analyzed near real-time, otherwise the systems implemented to analyze such data rapidly become obsolete over time \cite{Tian2019,XIE2020102659}. In machine learning and data mining, this phenomenon is referred to as concept drift.

\item Absence of platforms to reproduce empirical results: one of the main issues of the anomaly detection in energy consumption is the absence of platforms for reproducing the results of existing solutions. This may hinder the performance comparison between existing algorithms and make it difficult to understand the state-of-the-art.

\item Most of the frameworks differentiate between normal or abnormal power observations in general through separating them into two principal classes (normal and abnormal) without further details. However, in real-world scenarios, there exist different kinds of anomalous consumption, e.g. anomalies due to excessive consumption of an appliance are different from those due to keeping a door of the refrigerator open or those due to the absence of the end-user, as it is demonstrated in \cite{Himeur2020IJIS-AD}. In this line, without providing the end-user with the nature of anomalies and their sources, it is very difficult to trigger a behavioral change and promote energy saving. 
\end{itemize}

\subsection{Market drivers and barriers}
The frameworks reviewed in this article show that the anomaly detection topic is a promising strategy for a large number of services and applications in the energy field.
On the other hand, it is worth noting that the building energy monitoring market in general, comprises a multi-billion USD global opportunity. This market appears to be
growing at a robust rate, in which the anomaly detection takes a significant part \cite{AKHAVANHEJAZI201891}. The decision-making of energy saving systems in buildings depends on data, however, with the wide use of sub-meters and smart sensors, the data produced is very huge which can frequently provoke the loss or misunderstanding of relevant information \cite{himeur2020marketability}. Various active energy companies and utilities actually involved in providing anomaly detection and energy monitoring solutions, markedly illustrate the increased importance of this technology to promote energy efficiency. Table \ref{AnomalyProducts} summarizes a set of commercial anomaly detection of energy and energy management solutions developed by different companies, which are used for different kinds of buildings. Specifically, it provides a description of each solution, company name, frequency of energy monitoring and anomaly detection (real-time or near real-time), country and targeted building environments.

\begin{table} [t!]

\caption{A summary of existing AI-based energy monitoring and anomaly detection commercial solutions for buildings.}
\label{AnomalyProducts}
\begin{center}

\begin{tabular}{lllll}
\hline
{\scriptsize Product} & {\scriptsize Manufacturer} & {\scriptsize Description%
} & {\scriptsize Country} & {\scriptsize Implementation } \\ 
&  &  &  & {\scriptsize environment} \\ \hline
{\scriptsize Enetics SPEED \cite{Enetics2020}} & {\scriptsize Enetics} & 
{\scriptsize Faulty appliances identification and abnormal } & {\scriptsize %
USA} & {\scriptsize Public and domestic } \\ 
&  & {\scriptsize consumption detection} &  & {\scriptsize buildings} \\ 
{\scriptsize InBetween \cite{Ecoisme2020}} & {\scriptsize Ecoisme} & 
{\scriptsize Connected to the to the local Wi-Fi network and provides} & 
{\scriptsize Poland} & {\scriptsize Households} \\ 
&  & {\scriptsize consumption statistics via mobile app} &  &  \\ 
{\scriptsize Informetis \cite{Informetis2020} } & {\scriptsize Informetis} & 
{\scriptsize Near real-time energy monitoring and analysis using IoT } & 
{\scriptsize Japan} & {\scriptsize Households} \\ 
&  & {\scriptsize and big data mining technology} &  &  \\ 
{\scriptsize Verv Energy \cite{Verv2020}} & {\scriptsize Verv Energy} & 
{\scriptsize Real-time electricity consumption monitoring with iOS } & 
{\scriptsize UK} & {\scriptsize Households} \\ 
&  & {\scriptsize and \ Android app} &  &  \\ 
{\scriptsize Neurio \cite{Neurio2019}} & {\scriptsize Neurio} & {\scriptsize %
Real-time anomaly detection and notification } & {\scriptsize Canada} & 
{\scriptsize Households} \\ 
&  & {\scriptsize (appliances $\geq $ 400 W)} &  &  \\ 
{\scriptsize WiBeee HOME \cite{WiBeee2020}} & {\scriptsize WiBeee} & 
{\scriptsize Real-time consumption visualization, anomaly detection } & 
{\scriptsize Spain} & {\scriptsize Households} \\ 
&  & {\scriptsize using cloud and energy saving recommendation} &  &  \\ 
{\scriptsize Smart Impulse \cite{SmartImpulse}} & {\scriptsize Smart Impulse}
& {\scriptsize Building's energy consumption identification by end-use } & 
{\scriptsize France} & {\scriptsize Public buildings} \\ 
&  & {\scriptsize (lighting, IT, heating, etc.) and anomaly detection} &  & 
\\ 
{\scriptsize Verdigris \cite{Verdigris2020}} & {\scriptsize Verdigris} & 
{\scriptsize Energy consumption monitoring and real-time fault } & 
{\scriptsize USA} & {\scriptsize Industrial and commercial } \\ 
&  & {\scriptsize detection} &  & {\scriptsize buildings} \\ 
{\scriptsize Voltaware \cite{voltaware2020}} & {\scriptsize Voltaware} & 
{\scriptsize Real-time energy monitoring\ and anomaly detection using } & 
{\scriptsize UK} & {\scriptsize Commercial, industrial } \\ 
&  & {\scriptsize load disaggregation and tailored recommendations} &  & 
{\scriptsize and domestic buildings} \\ 
{\scriptsize HOMEpulse \cite{HOMEpulse2020}} & {\scriptsize HOMEpulse} & 
{\scriptsize Real-time energy disaggregation and anomaly detection } & 
{\scriptsize France} & {\scriptsize Households} \\ 
&  & {\scriptsize (1-10 sec \ sampling rate)} &  &  \\ 
{\scriptsize Hive Starter Pack \cite{AlertMe2020}} & {\scriptsize AlertMe} & 
{\scriptsize Electricity monitoring, appliances control using a mobile app} & 
{\scriptsize UK} & {\scriptsize Households} \\ 
{\scriptsize DiG Energy \cite{DigEnergy2020} } & {\scriptsize Intelen} & 
{\scriptsize Near real-time consumption monitoring, anomaly detection} & 
{\scriptsize USA} & {\scriptsize Commercial and domestic} \\ 
&  & {\scriptsize End-users education about energy efficient practices} &  & 
{\scriptsize buildings} \\ 
{\scriptsize Hark \cite{Hark2020}} & {\scriptsize Harksys} & {\scriptsize %
Real-time anomaly detection and cost saving through } & {\scriptsize UK} & 
{\scriptsize Residential and public } \\ 
&  &  &  & {\scriptsize buildings} \\ 
{\scriptsize EnerTalk \cite{enertalk2020}} & {\scriptsize ENCORED} & 
{\scriptsize Energy disaggregation, prediction and abnormal } & {\scriptsize %
Korea} & {\scriptsize Commercial and domestic } \\ 
&  & {\scriptsize usage detection} &  & {\scriptsize buildings} \\ \hline
\end{tabular}

\end{center}
\end{table}

In spite of the availability of the aforementioned solutions, different issues still require answers before enabling a widespread deployment of the anomaly detection technology in the energy industry.
First and foremost, anomaly detection solutions should demonstrate that they could provide the scalability, speed and privacy preservation needed for the considered application scenarios. Research efforts on distributed consensus algorithms, which are crucial to achieving these objectives, are still ongoing, however a solution that combines all desired characteristics cannot yet be achieved without significant trade-offs \cite{ANDONI2019143}.
Albeit anomaly detection systems could be installed using existing electric infrastructures, another crucial issue of these systems is that they have actually high implementation costs. Most of the solutions are built upon the latest machine learning methods, which require high-performance computing resources, e.g. using cloud platforms. Therefore, this slows down the commercialization of these solutions. 
Moreover, resistance to security attacks resulting from unintentionally inappropriate system development or theft attacks are not seriously addressed in most of the energy consumption anomaly detection solutions.

\section{Current trends and new perspectives}   \label{sec4}
\begin{figure*}[!t]
\centering
\includegraphics[width=\columnwidth]{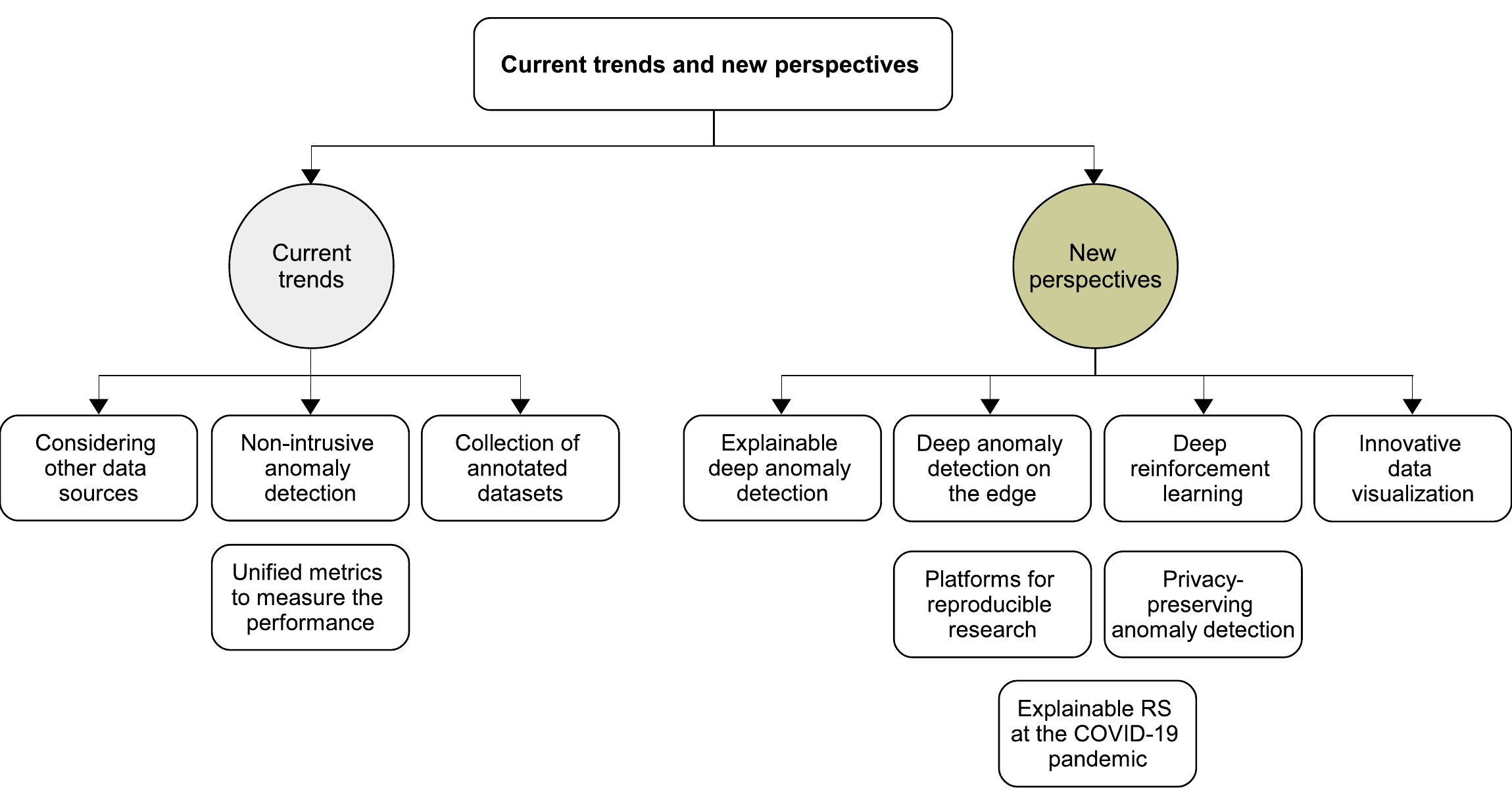}
\caption{List of current trends and new perspective of anomaly detection in energy consumption.}
\label{Trends} 
\end{figure*}

After reviewing anomaly detection frameworks, discussing their limitation and drawbacks, and describing important findings, it is of utmost importance to describe the current trends of this niche and derive the new perspectives that could be targeted. This aids the anomaly detection community in understanding the current challenges and future opportunities to improve the anomaly detection technology of energy consumption in buildings. Fig. \ref{Trends} summarizes the current trends and new perspectives that are identified in this framework.

\subsection{Current trends} \label{sec4-1}

Anomaly detection in energy consumption presents various challenges, which are mainly domain-specific. For instance, there is not a unique definition of normal versus anomalous consumption and there is inexplicit frontiers that separate normal and anomalous behaviors. Moreover, there is an absence of ground-truth data and unified metrics that could be deployed to evaluate the performance of anomaly detection algorithms. In addition, other data sources could result in triggering non conventional energy consumption anomalies, such as: presence/absence of end-users, opening of windows/doors when some specific appliances are on. To that end, this section discusses a set of current trends that should be considered to enhance the anomaly detection technology for energy saving applications.

\subsubsection{Considering other data sources}
In traditional anomaly detection schemes deployed for energy consumption, the anomalies are generally detected using only power consumption data gleaned from the main circuit or from individual devices, without paying any attention to other factors that can affect the consumption. However, in order to conduct an accurate anomaly detection, all the data that impact power consumption should be gleaned and stored along with energy consumption patterns. Following, anomaly detection algorithms should be built with reference to all these data, which can be summarized as follows: 

\vskip2mm

\noindent{\textbf{D1. Appliance parameters:}} each appliance has specific parameter settings that are responsible for its well functioning, such as the minimum standby consumption, maximum standby consumption and maximum operation time. These parameters are important to define normal and abnormal consumption of appliances and further to detect whether an appliance is working perfectly or is faulty.

\vskip2mm

\noindent{\textbf{D2. Occupancy patterns:}} the presence or absence of end-users could highly affect energy usage and results in some anomalous consumption behaviors that are not directly linked to excessive consumption of appliances. For example, turning on an air conditioner, television, fan or desktop when end-users are absent should be considered as an abnormal consumption behavior. To that end, recording occupancy data allows detecting unconventional anomalous consumption behaviors. 

\vskip2mm

\noindent{\textbf{D3. Ambient conditions:}} energy consumption could be extremely impacted by indoor conditions, such as the temperature, humidity and luminosity since the operation of some appliances depends mainly on these factors (e.g. air conditioners, heating systems, fans, light lamps, etc.). Therefore, collecting this kind of data aids in capturing abnormal energy consumption.

\vskip2mm

\subsubsection{Non-intrusive anomaly detection}
Starting from the advantage of NILM as a good alternative to sub-metering for collecting itemized billing, its use for detecting appliance-specific anomalies is very appreciated. Specifically, using NILM will remove the need to install individual sub-meters for each appliance and hence helps in significantly reducing the cost of anomaly detection solutions \cite{Rashid8537813,Himeur2020IJIS-NILM}. The use of NILM to detect abnormal consumption results in the development of a new kind of non-intrusive anomaly detection systems \cite{himeur2020smart}. In \cite{RASHID2019796,Rashid2017}, the authors have attempted to investigate if device-specific consumption fingerprints detected using NILM could be utilized directly to identify anomalous consumption behaviors and to what extent this could impact the accuracy of the identification. Accordingly, even though the performance of NILM to identify abnormal consumption is not yet as accurate as using sub-metering feedback, its performance could be further improved to allow a robust identification of faulty behavior. Moving forward, more effort should be put in this direction to develop non-intrusive anomaly detection of sufficient fidelity without the need to install additional sub-meters \cite{Rashid8683792,Himeur2020icpr}.

\subsubsection{Collection of annotated datasets}
\begin{figure*}[!t]
\centering
\includegraphics[width=0.9\columnwidth]{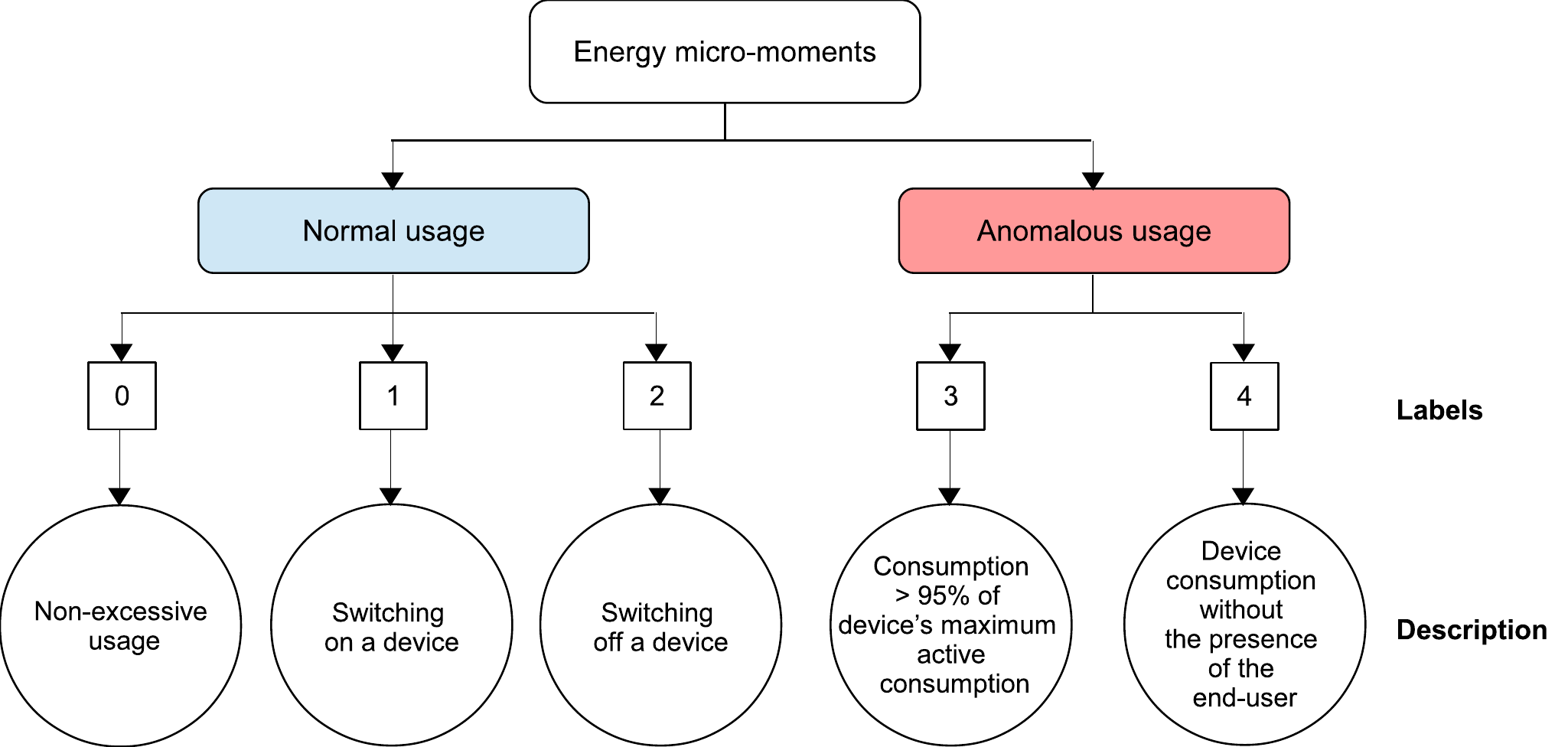}
\caption{Micro-moments assumption and labeling used in \cite{HimeurCOGN2020} to cluster normal and abnormal energy consumption data using a rule-based algorithm.}
\label{figMicroDesc} 
\end{figure*}

As mentioned previously, the absence of annotated datasets impedes the development of power anomaly detection solutions. To that end, greater effort should be put to collect and annotate power consumption datasets at different building environments (households, workplaces, public buildings, and industrial buildings), and further to share them publicly. This can help researchers to speed up the process of testing and validating their algorithms. In this context, the authors in \cite{HimeurCOGN2020} launch two new datasets for anomaly detection. The former, called Qatar university dataset (QUD) is collected in an energy lab and offers the consumption of four appliance categories along with the occupancy patterns for a period of three months. While the latter, named power consumption simulated dataset (PCSiD), produces consumption fingerprints of six devices and occupancy data for a period of two years. Both datasets provide power consumption footprints with their associated labels, where the overall data is split into five consumption classes. Three of them represent normal consumption classes, they are called \enquote{good consumption}, \enquote{turn on device} and \enquote{turn off device}, while the two remaining classes refer to anomalous consumption groups, which are defined as \enquote{excessive consumption} and \enquote{consumption while outside}. Fig. \ref{figMicroDesc} resumes the assumption and labeling process of micro-moment classes, which is applied in QUD and PCSiD \footnote{Both QUD and PCSiD datasets could be accessed via \url{http://em3.qu.edu.qa/}}.

\subsubsection{Unified metrics to measure the performance}
In addition to what has been been presented and based on analyzing the state-of-the-art, it is worth mentioning that there is no unified metrics and schemes to evaluate the performance of the anomaly detection algorithms. By contrast, a fair comparison between different anomaly detection approaches should be conducted using an ensemble of standard metrics, and should be performed under the same conditions, e.g. using the same dataset including appliance fingerprints collected at the same sampling rate \cite{Gaur8709671}. 

\subsection{New perspectives} \label{sec4-2}
Recently, governments, end-users, utility companies and energy providers pay a significant interest to the anomaly detection technology as a sustainable solution that could help in achieving the energy efficiency targets. In this section, we provide a general overview of new perspectives in anomaly detection in energy consumption.

\subsubsection{Explainable deep anomaly detection}
Deep learning based based anomaly detection solutions receive an increasing attention in current frameworks. However, despite their good performance, the black-box property of deep learning models represents a disadvantage in practical implementation \cite{Amarasinghe8430788}. Particularly, in energy consumption anomaly detection schemes, explanations of abnormalities detected using deep learning are critical. To that end, developing deep learning based abnormality detection techniques providing explanations why a power consumption observation/event is abnormal supports end-users/experts in focusing their investigations on the very crucial abnormalities and can boost their trust in the adopted solutions \cite{kauffmann2020clever,antwarg2019explaining}.  

For instance, one important orientation could be through developing a novel generation of explainable deep one-class learning models to effectively detect different kinds of energy consumption anomalies \cite{liznerski2020explainable}. Specifically, this category of models helps in (i) learning a mapping to concentrate normal consumption observations in a feature space, (ii) pushing abnormal patterns to be mapped away, and (iii) providing appropriate explanations for the anomalies detected, or more exactly, a human-readable prescription presenting helpful information on the causes that have led to the anomaly. Moreover, this enables generating tailored recommendations endorsing end-users to reducing their wasted energy and energy providers to detecting non-technical losses through the use of explainable recommender systems (RS) \cite{Sardianos2020GreenCom}.

\subsubsection{Deep anomaly detection on the edge}
Deep learning is one of the promising solutions to implement powerful anomaly detection solutions, however, a couple of years ago, it had been pretended that deep learning could just be implemented on high-end computing platforms, while the training/inference is conducted at the edge and carried out by edge servers, gateways or data centers. It had been a legitimate presumption at that period since the tendency was through the distribution of computing resources among the clouds and the edge serves. However, this situation is changed completely currently owing to recent R\&D achievements performed by academic and industrial partners \cite{10.1145/3287624.3287716}. Accordingly, the alternative considers the use of novel microcontrollers that include integrated machine learning accelerators. This could bring machine learning and specifically deep learning to the edge devices. The latter could not just execute machine learning algorithms, but they do that while consuming very low power and they need to connect to cloud just if required. Overall, this kind of microcontroller with embedded machine learning accelerators provides promising opportunities to offering computation capability for energy sub-meters and sensors collecting ambient conditions (i.e. temperature, humidity and luminosity), which gather data to enable various IoT applications \cite{sayed2021endorsing}.
  
\begin{figure*}[!t]
\centering
\includegraphics[width=\columnwidth]{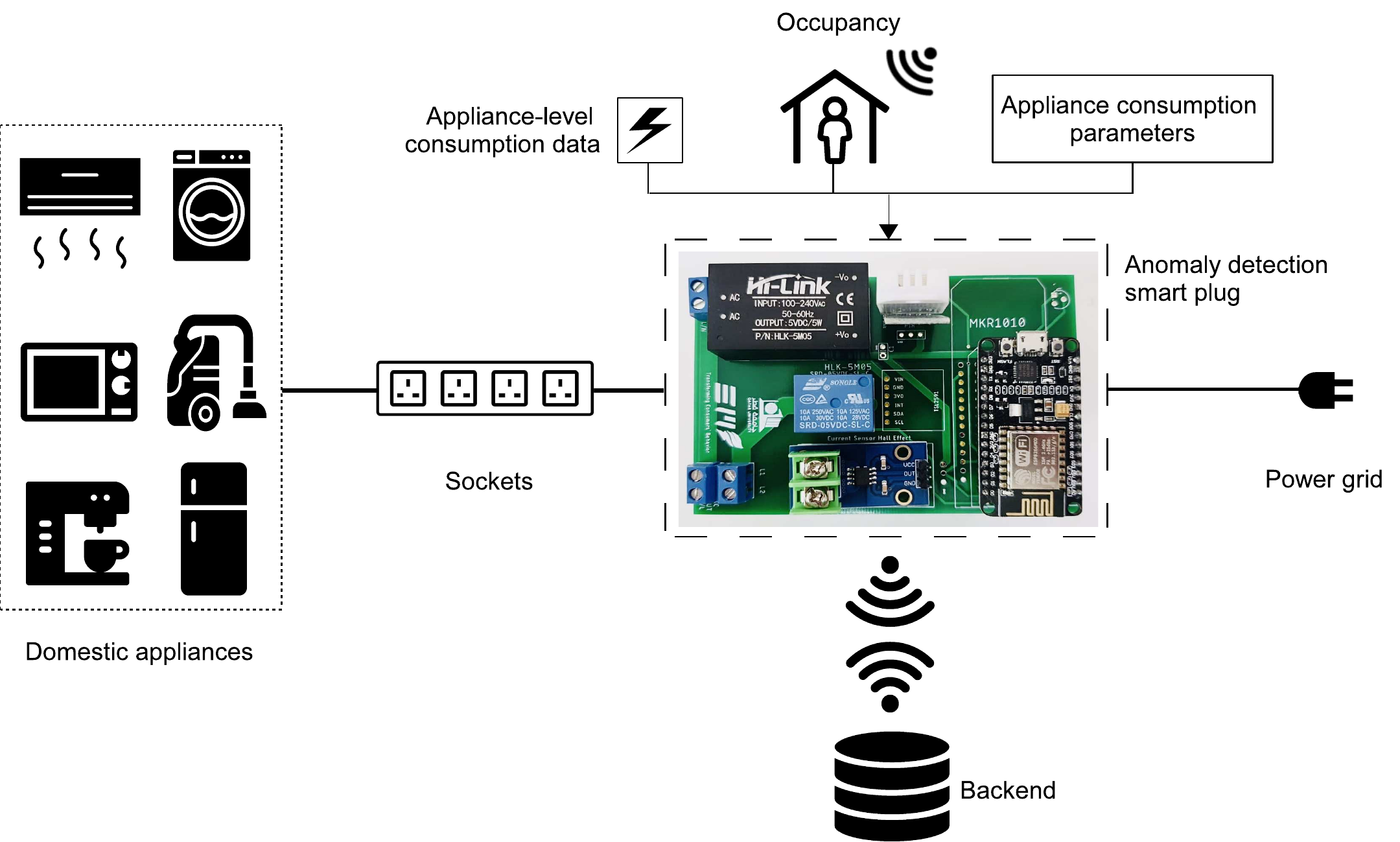}
\caption{Example of an edge-based anomaly detection solution used to develop a novel smart plug in the (EM)$^3$ framework.}
\label{fig07} 
\end{figure*}

On the other side, the edge is widely regarded as the furthest point in any IoT network that could be an advanced gateway (or edge server). Furthermore, it terminates at the sub-meters/sensors near the end-user. Thus, placing more analytical power near the end-user has become rational, where microcontrollers could be very convenient. Explicitly, this allows the inference and eventually the training, to be performed on tiny and resource-constrained low-power devices, instead of the large computing platforms (e.g. desktops, workstations, etc.) or the cloud. It is worth noting that to implement deep learning models, their size needs to be reduced in order to adapt the moderate computing, storage, and bandwidth resources of such devices, while maintaining the essential functionality and accuracy. Fig. \ref{fig07} illustrates an example of the anomaly detection solution embedded on a microcontroller based smart plug, which is under development in the (EM)$^3$ project \cite{Alsalemi8959214}.

\subsubsection{Deep reinforcement learning}
Reinforcement learning is a promising topic of AI that has received a significant attention recently. Its concept is related to comprehending the human decision-making procedure before developing algorithms for enabling agents to determine the proper anomaly behavior using trial-and-error in parallel with the reception of feedback form of reward power consumption signals \cite{Aberkane8966795}. In this regard, deep reinforcement learning (DRL) is then proposed as a merge of deep learning and reinforcement learning to detect more complex consumption anomalies. Detecting such abnormalities involves handling high-dimensional consumption patterns and environmental conditions, uncertainties of the agent's observations and sparse reward power consumption signatures. DRL techniques have been proposed lately to resolve a broad variety of issues, including detecting abnormalities video surveillance, traffic management and anomaly detection \cite{Shabestary8569549,app10114011}, communication and networking \cite{Luong8714026} and energy consumption prediction \cite{LIU2020109675}.

Overall, DRL shows promising opportunities to resolve effectively the problem of energy consumption anomaly detection since the latter is considered as a decision-making task. Following, an agent is designed to learn from the consumption and environmental data via a continuous interaction with them and reception of rewards for detected anomalies, i.e. the process is similar to the natural human learning via their experiences. 

\subsubsection{Multimodal anomaly visualization}
As explained previously, the capability to interpreting anomalous and normal power consumption behaviors is of utmost importance since the essential intrinsic challenges in the abnormality detection issue are mainly related to (i) the absence of obvious boundaries between anomalous and normal consumption observations and (ii) the complexity to obtain annotated power consumption datasets to train and verify developed solutions. To that end, the knowledge and experience of human experts are highly esteemed to judge the consumption scenarios. A subjective, comprehensive and interactive visualization of power consumption patterns and resulted analytic is hence greatly helpful to support the interpretation and facilitate an optimal decision-making. In this context, great attention has been devoted recently to using innovative visualization tools and visual analysis methods to detect anomalous data in other research fields, such as the spreading of rumors on social media \cite{ZHao6876013} and user behavior \cite{shi2019visual,Guo9005687}.

In this regard, using visualization and interactivity for detecting anomalous power consumption behaviors and supporting end-users' interpretability and interactivity represent a promising research direction, especially to understand sense-making of anomalous consumption footprints and explain why an anomaly occurs. For instance, novel visualization plots are designed in the (EM)$^3$ framework to portray anomalous consumption patterns using a scatter plot, in  which two kind of anomalies, i.e. \enquote{excessive consumption} and \enquote{consumption without the presence of the end-user} along with normal data are traced over the day time.

\begin{figure*}[!t]
\centering
\includegraphics[width=\columnwidth]{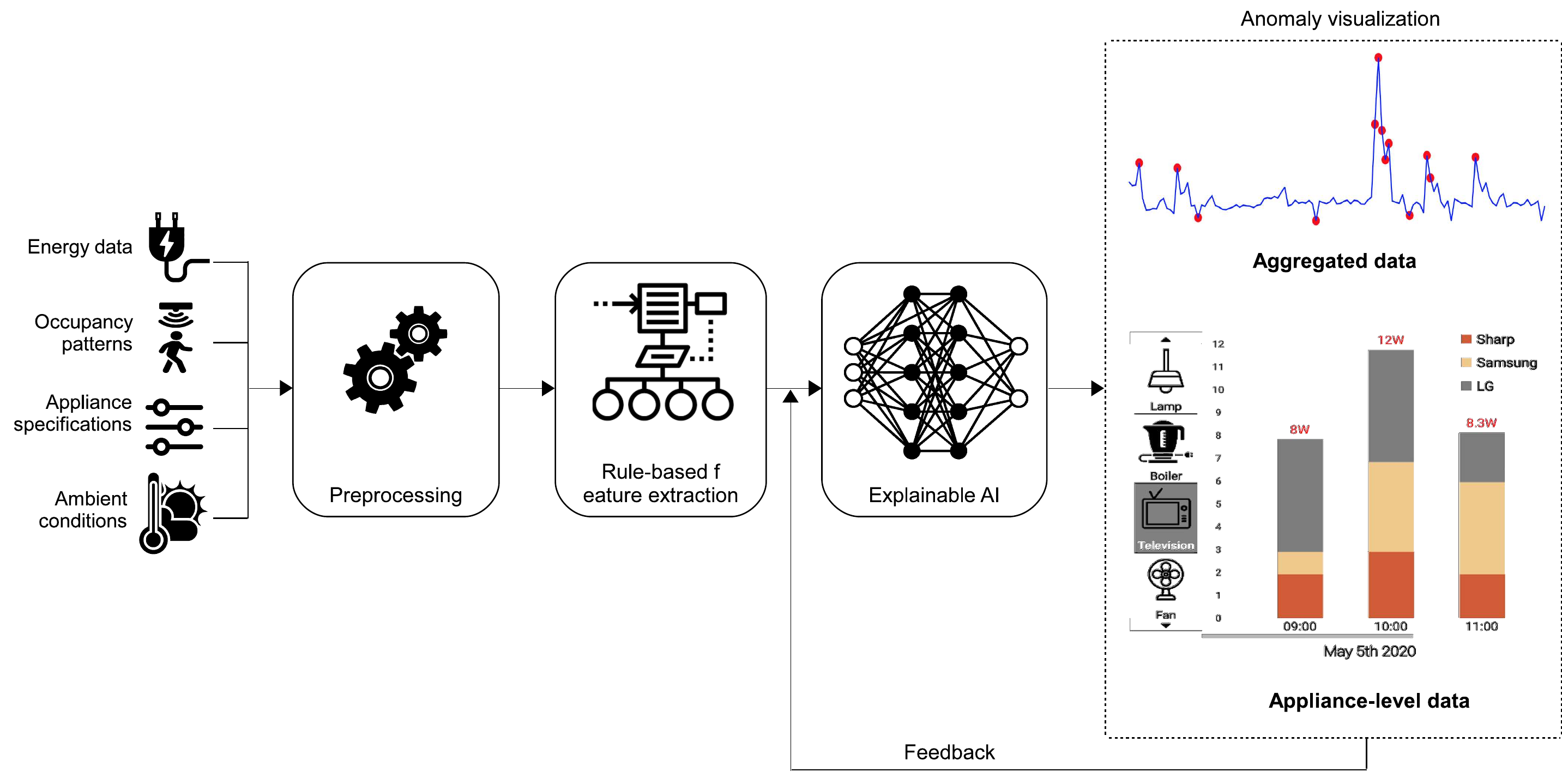}
\caption{A novel architecture of visualization based anomaly detection in energy consumption to (i) improve the accuracy of detected anomalies, and (ii) help end-users in comprehending their energy consumption footprints.}
\label{StackedBar} 
\end{figure*}

Furthermore, another notable visualization plot developed in (EM)$^3$, which could provide end-users with consumption analytics and anomaly detection capabilities at an appliance-level is the stacked bar \cite{AymanSETA2020}. It enables to select devices and stack various models of the same device altogether (e.g. televisions from distinct brands). Visualizing multi-level power consumption could help end-users in effectively detecting anomalies and faulty devices, and hence could allow them to perform better decision-making towards reducing wasted energy \cite{AymanGPECOM2020}. Fig. \ref{StackedBar} portrays our perception of a multimodal visualization based anomaly detection of energy consumption, in which visualization feedback (either at the aggregated level or the appliance level) could be used to improve the accuracy of anomaly detection.


\subsubsection{Platforms for reproducible research}
Despite the advance achieved in developing anomaly detection methods for energy consumption, principally, three aspects affect reproducibility, and thus a fair and experimental comparison of anomaly detection algorithms: (i) it is difficult to evaluate the generality anomaly detection techniques as most of the frameworks are generally assessed on a unique dataset, (ii) there is an absence of frameworks comparing existing solutions under the same conditions, because of the lack of available open-source anomaly detection datasets and (iii) distinct and standardized assessment criteria are used in the state-of-the-art with regard to the considered scenario \cite{10.1145/3360322.3360844}. 

To overcome this issue, there is an urgent need to release an open source anomaly detection toolkit, which includes challenging energy consumption datasets and existing anomaly detection algorithms. This will allow a fair and easy comparison of anomaly detection algorithms in a reproducible manner. Furthermore, this will support the foundations for future anomaly detection  competitions \cite{10.1145/2602044.2602051}.

\subsubsection{Privacy-preserving machine learning}
The wide use of machine learning methods for anomaly detection in energy consumption is actually limited by the lack of open-access anomaly detection datasets to train and validate algorithms, due to strict legal and ethical requirements to protect end-user privacy. Aiming at preserving end-user privacy  while promoting scientific research while using power consumption datasets, implementing novel approaches for federated, secure and privacy-preserving machine learning is an urgent need. In this context, removing private information (anonymization) and replacing of vulnerable inputs with artificially produced ones while permitting a reattribution based on a look up table (pseudonymization) are among the solutions that could be targeted \cite{Kaissis2020}. Furthermore, using federated machine learning, which helps in training algorithms over various decentralized edge-devices/servers holding local power consumption patterns without sharing them seems very promising for anomaly detection in energy consumption \cite{Hao8761267}.

\subsubsection{Explainable RS and the COVID-19 pandemic}
Power consumption in buildings has been completely changed in the COVID-19 pandemic due to the constraints on movement. This has widely triggered remote working and e-learning, and hence has shifted activities and energy usage to domestic residents \cite{MADURAIELAVARASAN2020115739}. Therefore, the need for smart solutions to detect energy consumption anomalies with reference to the actual situation and other changes that could be occurred at any time is a current challenge \cite{sardianos2020smart}. To that end, the use of RS for supporting human decision making has recently received increased interest \cite{Nilashi2020,Budd2020}. However, with the aim of increasing the end-user trust and improving the acceptance of the generated recommendations, these systems should provide explanations \cite{alsalemi2020micro}.

\begin{figure*}[!t]
\centering
\includegraphics[width=\columnwidth]{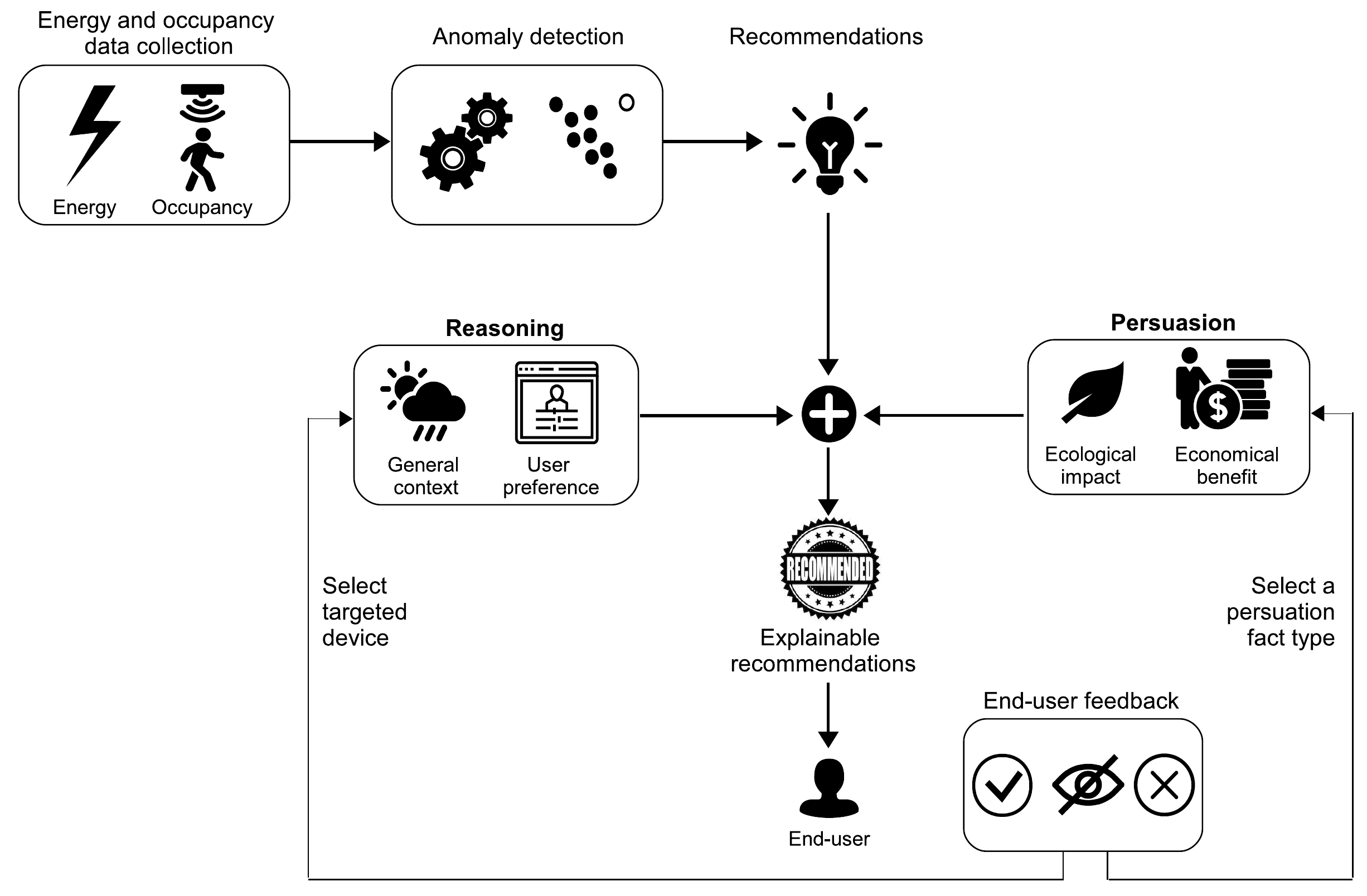}
\caption{Example of the flowchart of an energy saving system based on the combination of anomaly detection and RS, in which the output of the anomaly detection module serves as an input for the RS to help end-users in correcting their energy consumption behaviors.}
\label{RecSys} 
\end{figure*}

In this context, developing mechanisms for explainable and persuasive energy consumption recommendations that could be tailored based on the end-user preferences, habits and current circumstances will promptly reduce wasted energy and promote energy saving \cite{himeur2020survey}. Specifically, the explanations could justify the reasons for recommending each energy efficiency act \cite{INR-066}. On the other hand, the persuasiveness of fact-based explanations could be improved using persuasive and incentive aspects, such as emphasizing ecological impacts and economical saving benefits. Fig. \ref{RecSys} illustrates a general flowchart of an explainable energy RS proposed in the (EM)$^{3}$ framework \cite{Sardianos2020IJIS-ERS}. Moreover, it is worth noting that explainable RS are much appropriate to unexpected energy consumption situations (e.g. the COVID-19 pandemic) since the recommendations could be generated in real-time in addition to providing the end-user with more details (using contextual data) on each recommended action to increase its acceptance \cite{sardianos2020data}.


\section{Conclusion} \label{sec5}
In this article, a systemic and technically-informed survey of anomaly detection methods in building energy consumption has been presented. A taxonomy that classifies these approaches with reference to different aspects has been presented, such as artificial intelligence models, application scenarios, detection level and computing platforms. To conclude, anomaly detection strategies can evidently benefit energy saving systems, energy providers, end-users and governments via reducing wasted consumption and energy costs. Specifically, they provide insight information on abnormal consumption behavior, anomalous appliances, non-technical loss and electricity theft cyberattacks, but most significantly, anomaly detection systems offer smart and powerful solutions for promoting energy saving. They also play a major role in the energy monitoring market. 

We have showed that the majority of anomaly detection solutions in energy consumption are still in their nascence. To promote their widespread utilization and maturity, a set of challenges and limitations should be overcome, among them the lack of annotated datasets, absence of the reproducibility platforms, and the lack of standard metrics to assess the performance developed solutions. On the other hand, energy consumption is impacted by other factors such as, occupancy (presence/absence of the end-user), ambient conditions, outdoor temperatures and end-user's preferences. Therefore, it is of utmost importance to consider these data to develop powerful and reliable anomaly detection models, which could detect more advance kinds of abnormal energy usage.
All in all, a significant research effort should be made in the near future to confront the aforementioned issues and improve the quality of anomaly detection systems.

In addition, further investigations are still ongoing in future directions, which could permit developing power anomaly detection systems in terms of the scalability, decentralization, low power consumption, easy implementation and privacy preservation. Finally, we believe that more research contributions, projects and collaborations with industrial partners should be performed to help anomaly detection technology reach its entire potential, proving its commercial feasibility and facilitating its mainstream adoption in residential buildings.

\section*{Acknowledgements}\label{acknowledgements}
 This paper was made possible by National Priorities Research Program (NPRP) grant No. 10-0130-170288 from the Qatar National Research Fund (a member of Qatar Foundation). The statements made herein are solely the responsibility of the authors.


\end{document}